\documentclass[conference,compsoc]{IEEEtran}
\IEEEoverridecommandlockouts
\usepackage{tikz}
\usepackage{amsmath}

\usepackage{pifont}
\usepackage{filecontents}
\usepackage{adjustbox}
\usepackage{color}
\usepackage{tabularray} 
\usepackage{multirow}
\usepackage{graphicx}
\usepackage{xcolor}
\usepackage{tabularx}
\usepackage{booktabs}
\usepackage{hyperref}
\usepackage{amsmath}
\usepackage{booktabs,caption}
\usepackage[flushleft]{threeparttable}
\usepackage{makecell}
\usepackage{algorithm}
\usepackage{algorithmicx}
\usepackage{algpseudocode}
\usepackage{balance}
\usepackage{tabu}
\usepackage{caption}
\usepackage{subcaption}
\usepackage{pifont}
\usepackage{tikz}
\usepackage{xspace}
\usepackage{amssymb}
\usepackage{tcolorbox}
\newcommand{\hmark}{{\ding{51}}\textsuperscript{\kern-0.52em\scriptsize\ding{55}}}
\newcommand{\system}{\textsc{ProvFusion}}
\usepackage{booktabs}
\usepackage{multirow}
\usepackage{makecell}
\usepackage{siunitx}
\usepackage{wrapfig}
\usepackage[normalem]{ulem}
\usepackage{colortbl}

\usepackage{xcolor}
\definecolor{diffblue}{RGB}{0,0,180}
\newcommand{\rev}[1]{{#1}}
\usepackage{fancyhdr}
\usepackage{lastpage}
\newcommand{\arxivacceptednotice}{%
\thispagestyle{fancy}%
\fancyhf{}%
\fancyfoot[C]{%
\parbox{0.95\textwidth}{\centering\footnotesize
© 2026 IEEE. Personal use of this material is permitted. Permission from IEEE must be obtained for all other uses, in any current or future media, including reprinting/republishing this material for advertising or promotional purposes, creating new collective works, for resale or redistribution to servers or lists, or reuse of any copyrighted component of this work in other works.
\\
Accepted for publication in the 2026 IEEE Symposium on Security and Privacy (IEEE S\&P 2026).
}}
\renewcommand{\headrulewidth}{0pt}
\renewcommand{\footrulewidth}{0pt}
}

\ifCLASSOPTIONcompsoc
  \usepackage[nocompress]{cite}
\else
  \usepackage{cite}
\fi

\ifCLASSINFOpdf
\else
\fi
\begin{document}
\title{Beyond Nodes vs. Edges: A Multi-View Fusion Framework for Provenance-Based Intrusion Detection}

\author{
Fan Yang\textsuperscript{\dag},
Binyan Xu\textsuperscript{\dag},
Di Tang\textsuperscript{\ddag *}\thanks{* Corresponding authors},
Kehuan Zhang\textsuperscript{\dag *}\\[2pt]
{\itshape
\textsuperscript{\dag}The Chinese University of Hong Kong \qquad
\textsuperscript{\ddag}Sun Yat-Sen University
}\\[2pt]
{\small\texttt{\{yf020, binyxu, khzhang\}@ie.cuhk.edu.hk \quad tangd9@mail.sysu.edu.cn}}
}

\maketitle
\arxivacceptednotice


% \fi
\IEEEpeerreviewmaketitle

\begin{abstract}
Provenance-based intrusion detection has emerged as a promising approach for analyzing complex attack behaviors through system-level provenance graphs. However, existing defense methods face an inherent granularity limitation. Node-centric detectors, which evaluate anomalies using entities’ attributes and local structural patterns, may misclassify benign behavioral changes or configuration modifications as suspicious. In contrast, edge-centric detectors, which focus more on interactions, may lack sufficient contextual awareness of the involved entities, leading to missed detections when compromised entities perform seemingly ordinary operations. These analytical biases highlight a persistent gap between node-centric and edge-centric analyses.

To mitigate this gap, we present \system{}, a multi-view detection framework that integrates anomaly signals from three distinct views (i.e., attribute, structure, and causality). The framework fuses heterogeneous anomaly signals through lightweight fusion schemes and determines the final anomaly decisions through a voting-based integration process, providing a more consistent and context-aware assessment of system behavior. This design enables \system{} to capture both entity-level deviations and interaction-level anomalies within a consistent analytic pipeline. Experiments on \rev{nine} widely used benchmark datasets demonstrate that \system{} achieves higher detection accuracy and lower false-positive rates than single node- and edge-centric baselines, maintaining stable performance across scenarios. Overall, the results suggest that our multi-view anomaly fusion together with voting-based decision aggregation offers a practical and effective direction for advancing provenance-based intrusion detection.
\end{abstract}


% \fi
% \IEEEpeerreviewmaketitle

% \vspace{-7pt}
\section{Introduction}
\vspace{-9pt}
Intrusion Detection Systems (IDSs)~\cite{nodoze,rapsheet,priotracker,sok,hercules} have long served as a cornerstone of system defense, particularly against sophisticated threats such as Advanced Persistent Threats (APTs)~\cite{apt}. Among them, provenance-based intrusion detection systems (PIDSs)~\cite{holmes,winowner} have received growing attention in recent years~\cite{shadewatcher,provdetector,unicorn,edgetorrent}. These systems construct provenance graphs from system audit logs, where nodes represent entities such as processes and files, and directed edges represent causal or information-flow interactions (e.g., read, write). By preserving the causal chains of system behavior, provenance graphs provide a structured and interpretable foundation for attack detection, forensic analysis, and damage assessment, thereby playing an increasingly indispensable role in modern IDSs.

With the development of Graph Neural Networks (GNNs)~\cite{powerfulgnn}, provenance-based detection has evolved from handcrafted or statistical rules toward learning-based analysis~\cite{altinisik2023provgsearchergraphrepresentationlearning,ProGrapher,provenance_survey}. GNNs integrate node/edge attributes with local topology via message passing~\cite{magic,flash}, producing contextualized representations that support fine-grained \emph{local} reasoning over dependencies within a $L$-hop neighborhood, where $L$ is the number of GNN layers.
These advances have improved PIDSs' capacity to model local dependencies and to score entities/edges individually, supporting a move \emph{from coarse-grained, subgraph-level alerts} to \emph{finer-grained attack localization}~\cite{akoglu2014graphbasedanomalydetectiondescription,Ma_2023}.

Despite these advances, existing fine-grained GNN-based PIDSs adopt a \emph{single} detection view—either node-centric~\cite{magic,flash,rcad,sigl,nodlink} or edge-centric~\cite{velox,orthrus,kairos}.
Node-centric methods assess whether an entity is anomalous based on attribute semantics and the structural context of its $L$-hop neighborhood; benign dependency changes (e.g., software or configuration updates) can shift representations away from the training distribution and inflate false positives.
Edge-centric methods evaluate individual interactions based on aggregated representations of the involved entities; high-frequency edges may be regarded as normal even when the involved entities are compromised, \emph{leading to missed detections} (e.g., a compromised process performing ordinary-looking actions that are classified as normal).
Thus, each detection view is informative but inherently limited, providing a partial view of system behavior.

The inherent limitations of node- and edge-centric detection views reflect a \textit{fundamental challenge}: the behavioral information encoded in provenance graphs is only partially utilized. Therefore, an effective intrusion detection system should integrate these views to cross-reference entity- and interaction-level anomalies, enabling a more context-aware assessment of system execution.

Achieving such integration, however, is non-trivial. Different detection views naturally produce heterogeneous anomaly scores, each emphasizing distinct behavioral factors and exhibiting different statistical scales. Naïve aggregation (e.g., summation or averaging) can miscalibrate each view’s contribution, and a single averaging rule cannot adapt to different operating contexts, leading to unstable or biased detection outcomes across datasets and environments.

To address the difficulty of integration, we develop \system{}, a unified provenance-based intrusion detection framework that embodies multi-view fusion with adaptive, voting-based detection. 

It operates at both the score and decision stages. At the score stage, heterogeneous anomaly scores from different detection views are integrated through a multi-view fusion paradigm that evaluates each system entity across seven \textit{dimensions of anomaly}, each intended to capture a distinct behavioral characteristic relevant to potential anomalies. At the decision stage, to ensure practicality and stability, we introduce a \textbf{voting-based adaptive detection mechanism} that makes the final anomaly decisions based on the vote result of multiple scoring dimensions. 
This two-stage design enhances the utilization of behavioral information while stably producing effective results.

We evaluate \system{} on six datasets from the DARPA Transparent Computing program (E3~\cite{darpa3program} and E5~\cite{darpa5})\rev{, and three host subsets of the DARPA OpTC dataset~\cite{darpaoptc}, which} are widely used in prior PIDS research~\cite{orthrus,velox,kairos,magic,flash}. \rev{On the TC datasets,} \system{} achieves higher detection quality—more true positives (TPs) with fewer false positives (FPs)—than SOTA node- and edge-centric baselines (on average, 24.8 TPs and 6.2 FPs vs. the second-best method’s 7.2 TPs and 136.5 FPs at the node level), and higher attack-level recall (detecting 13/14 attacks vs. 12/14 for the second best). \rev{On OpTC, \system{} achieves the best detection–FP trade-off (avg 7 TPs / 13 FPs), while node-centric baselines incur $10^4$--$10^5$ FPs and edge-centric baselines collapse to $< 3$ TPs.}
In addition, an in-depth ablation study shows that \system{} outperforms its single-view variants, further supporting the benefit of combining distinct detection views.

Our main contributions are as follows:

\noindent(1) \textbf{Characterizing the granularity biases.} We analyze the biases of node- and edge-centric detection view in GNN-based PIDSs, and provide an empirical and conceptual understanding of how these views capture distinct but incomplete aspects of provenance behavior.

\noindent(2) \textbf{A multi-view detection framework.} We present \system{}, a multi-view framework that unifies node- and edge-centric views within a single pipeline, combining three distinct detection views—attribute, structure, and causality—to capture diverse facets of system behavior.

\noindent(3) \textbf{A multi-dimensional fusion and voting-based detection mechanism.} We design a two-level fusion-and-detection process that first integrates heterogeneous anomaly scores through monotone fusion schemes, and then achieves stable detection through voting-based integration.

\noindent(4) \textbf{Empirical evaluation and reproducibility.} We implement \system{} and evaluate it on \rev{nine} DARPA \rev{TC and OpTC} datasets. The results demonstrate that \system{} achieves consistently high accuracy, lower false positive rates, and comparable efficiency across multiple datasets, supporting its practical applicability. 
To foster reproducibility, we will release the source code: (\rev{\url{https://github.com/Joney-Yf/ProvFusion}}) along with a manually refined ground-truth label set based on official DARPA reports to support future research and benchmarking.

\section{Motivation: Understanding How Different Detection Views Focus Differently}
\label{sec:motivation}
\vspace{-9pt}

Provenance-based intrusion detection systems (PIDSs) typically analyze system behavior from two complementary granularities: the \textit{state or structure} of system entities (nodes)~\cite{magic,flash,nodlink} and the \textit{interactions} (edges) between them ~\cite{velox,orthrus}.  
Both detection views have demonstrated effectiveness, yet each exhibits different \textbf{focus} that can lead to false alarms or missed detections.  
These differences reflect the intrinsic emphasis of each detection view rather than particular implementations.

To ensure representativeness, we \textit{reproduced several state-of-the-art systems under the standardized \textit{Velox} benchmark}~\cite{velox}.  
Specifically, we included MAGIC~\cite{magic}, NodLink~\cite{nodlink}, and FLASH~\cite{flash} as three \textit{node-centric} methods, together with Velox~\cite{velox}, Kairos~\cite{kairos} and Orthrus~\cite{orthrus} as three \textit{edge-centric} baselines.  
% All reproductions adhere to the official configurations and dataset splits of the public DARPA Transparent Computing dataset.
Detailed experimental setting is in Section~\S\ref{sec:compared_baseline} and all our analyses are based on results in Table~\ref{tab:main_results}.

\vspace{-9pt}
\subsection{Node-Centric Perspective and Challenges}
\vspace{-9pt}

Node-centric PIDSs evaluate whether a system entity (e.g., a process or file) deviates from normal patterns, based on its attributes and neighborhood structure, usually by leveraging Graph Neural Networks (GNNs)~\cite{gcn,gat,powerfulgnn}.  
Our reproductions suggest two recurring difficulties that limit their robustness.

\vspace{-9pt}
\subsubsection{Sensitivity to Benign Novelty}
MAGIC and NodLink frequently produce false positives when encountering benign but previously unseen behaviors (e.g., MAGIC and NodLink have over 90k false positives in THEIA-E3 dataset).  
Here, “novelty” refers to normal variations such as software updates, configuration changes, or new user activities.  
In our reproductions, entities introducing unseen attributes or structural patterns—such as a file with a new path—often received high anomaly scores despite being benign.  
This observation suggests that these node-centric models primarily focus on how unusual an entity and its structural patterns appear, which makes them sensitive to benign but previously unseen behaviors that are not necessarily malicious.

\vspace{-9pt}
\subsubsection{Objective–Security Misalignment in Type Prediction}
A similar tendency appears in FLASH, which predicts the type of each node.  
While this task captures categorical semantics, it does not necessarily align with security relevance.  
For instance, a compromised \texttt{process} may still exhibit process-like characteristics and thus be assigned a low anomaly score.  
In our reproductions, several malicious processes exhibited anomalous interactions that received high anomalous scores under edge-centric methods such as Orthrus. Yet their node-level anomaly scores remained low, suggesting that node-type prediction alone 
tends to evaluate local patterns rather than interaction abnormality, providing limited evidence of compromise.

\noindent\textbf{Summary:}  
Overall, node-centric analysis primarily focuses on an entity’s local patterns or structural consistency. 
This focus allows it to capture statistical irregularities but limits its awareness of whether such irregularities correspond to truly suspicious interactions in context.

\vspace{-9pt}
\subsection{Edge-Centric Perspective and Challenges}
\vspace{-9pt}

Edge-centric methods shift the analytical focus from entities to their interactions. The detection of anomalous interactions is often formulated as edge-type prediction~\cite{velox,kairos,orthrus}.  
Given the larger sample of edges and the more uniform distribution of edge types relative to node's structural patterns, this view tends to yield more precise results.
However, since it models interactions, it has limited awareness of the broader behavioral context in which the involved entities operate.

On one hand, edge-centric detectors may yield \textbf{false positives from rare-but-benign events}, where legitimate yet infrequent operations are assigned high anomaly scores due to their rarity.  
On the other hand, they may incur \textbf{false negatives from common-but-malicious events}, as attackers often exploit high-frequency system calls to blend into normal activity.  
In our reproductions, some attack interactions are structurally anomalous and detected by node-centric methods (e.g., MAGIC). Yet under the edge-centric view, they received low anomaly scores because their event types are common in benign traces.
These results suggest that assessing interaction-level plausibility alone may not fully capture the anomalous state of the participating entities.

\noindent\textbf{Summary:}  
Edge-centric analysis focuses on evaluating the plausibility of interactions between entities.  
While this view enables stable learning of interaction patterns, it provides limited awareness of the broader behavioral context in which those entities participate.

\vspace{-9pt}
\subsection{Visualization and Design Implications}
\vspace{-9pt}

We align the anomaly scores of different detection views on the same set of ground-truth entities to visualize how their focuses differ.
We take \textbf{MAGIC (node-centric)} and \textbf{Orthrus (edge-centric)} as representative examples. 

\begin{figure}[t]
    \centering
    \includegraphics[width=0.8\columnwidth]{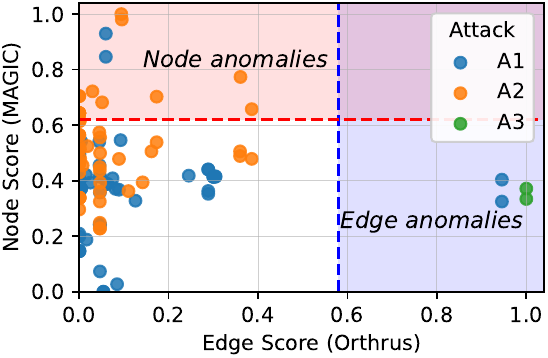}
    \vspace{-6pt}
    \caption{\textbf{Visualization of node- and edge-centric detection.}  
    No malicious nodes show high scores in both methods.}
    \label{fig:node_edge_tradeoff}
    \vspace{-12pt}
\end{figure}

As shown in Figure~\ref{fig:node_edge_tradeoff}, most detected malicious points fall into exactly one high-score region. Some lie above the red line but left of the blue line (flagged only by MAGIC); others lie to the right of the blue line but below the red line (flagged only by Orthrus). The upper-right quadrant—where both views would agree—is visibly empty. This pattern indicates that the two views capture different facets of abnormality and that either view alone provides incomplete attack coverage. 

Together, these findings motivate the multi-view framework introduced in Section~\ref{sec: framework},  
which integrates both views to form a more balanced and context-aware foundation for intrusion detection.

\begin{figure*}[t]
    \centering
    \includegraphics[width=0.95\textwidth]{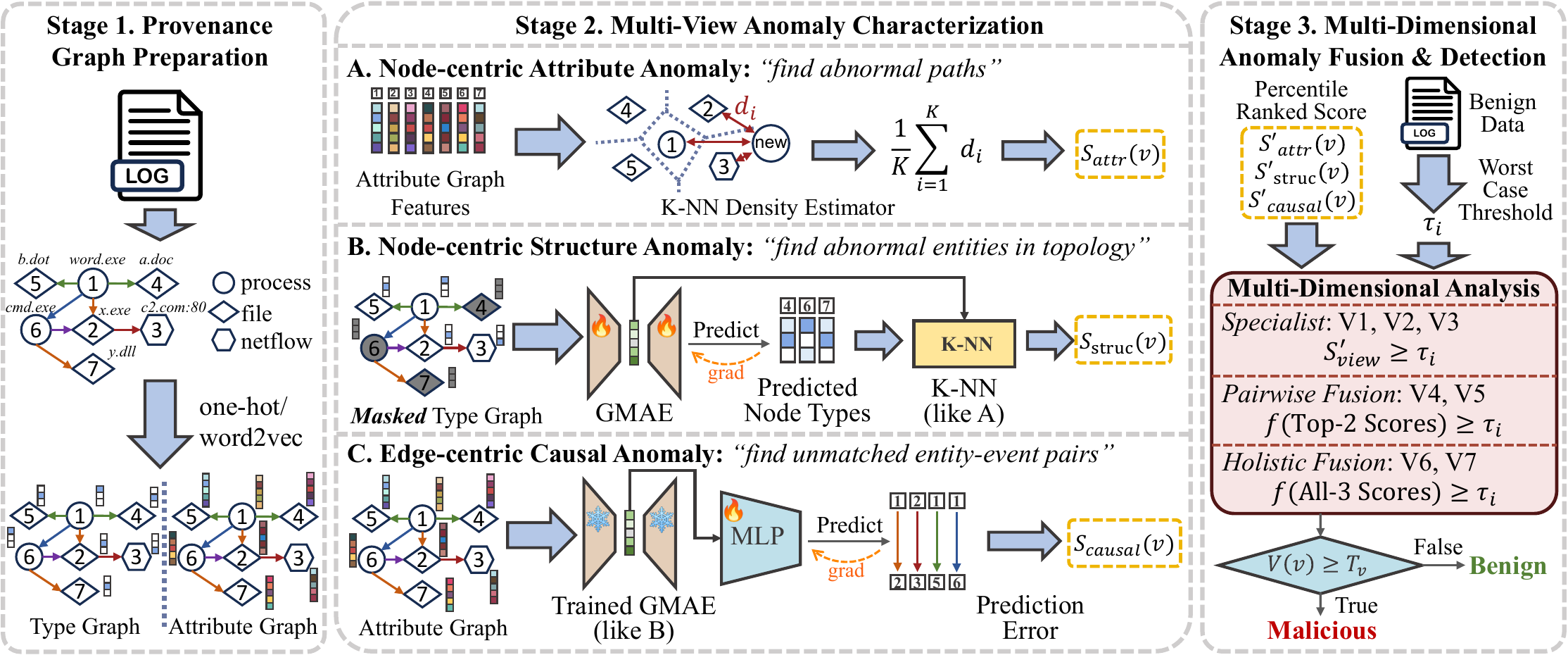}
    \vspace{-3pt}
    \caption{The overall architecture of \system{}. It takes raw system logs, prepares a provenance graph, characterizes node anomaly from three complementary views (Attribute, Structural, Causal), and finally fuses these scores to detect anomalies.}
    \vspace{-10pt}
    \label{fig:framework_overview}
\end{figure*}

\vspace{-9pt}
\section{Threat Model}
\label{sec:threat_model}
\vspace{-9pt}

Being consistent with prior work~\cite{kairos,flash,orthrus,velox,magic}, we consider adversaries whose objective is to compromise a host system and maintain unauthorized access or control over time. 
Our analysis focuses on activities captured by standard kernel-level monitoring frameworks~\cite{ETW,audit,Dtrace}, and therefore excludes threats that operate outside this scope (e.g., hardware-level side channels). 

We assume that data collection and model training occur in a trusted environment, as in prior studies~\cite{orthrus,kairos,velox}, thereby excluding data and model poisoning from our threat model. 
The Trusted Computing Base (TCB) includes the \system{} software, the provenance capture mechanism, and the underlying operating system. 
These components are assumed to be protected from direct compromise through system-hardening techniques described in prior research~\cite{custos,trustworthyProvenance}. 
Finally, we assume that the integrity of the provenance record is ensured by tamper-evident or append-only logging mechanisms~\cite{auditframeDefense,practicalwholeprocenance}, which make any modification attempts detectable.

\vspace{-9pt}
\section{\system's Framework}
\label{sec: framework}
\vspace{-9pt}
This section introduces \system{}, an anomaly-based intrusion detection system. As established in Section~\S\ref{sec:motivation}, methods that rely on a single view—either focusing on node states (\textit{node-centric}) or interactions (\textit{edge-centric})—exhibit distinct strengths and weaknesses. Our design goal is to leverage these distinct views to improve detection performance and reduce avoidable errors.

Based on this goal, we design \system{}: \textbf{a multi-view anomaly fusion and voting-based detection framework}. To avoid conflating signals, we treat attributes and structure as separate views (details in \S\ref{sec:multi_view_characterization}).
Specifically, \system{} characterizes each system entity from three distinct views:
(i) an \textbf{attribute view} that captures deviations in an entity’s intrinsic features,
(ii) a \textbf{structural view} that identifies abnormalities in the entity’s role and structure patterns, and
(iii) a \textbf{causal view} that evaluates the plausibility of interactions (i.e., edges) the entity participates in.
We quantify each view independently, fuse them across seven anomaly dimensions, and finalize decisions via a voting-based mechanism, yielding stable performance across heterogeneous score scales.

Figure~\ref{fig:framework_overview} illustrates the overall architecture. The framework has three stages:
(1) Provenance Graph Preparation (\S\ref{sec:graph_preparation}),
(2) Multi-view Anomaly Characterization (\S\ref{sec:multi_view_characterization}), and
(3) a Multi-Dimensional Anomaly Fusion and Voting-based Detection Framework (\S\ref{sec:fusion_detection}).

\vspace{-9pt}
\subsection{Provenance Graph Preparation}
\vspace{-9pt}
\label{sec:graph_preparation}
We transform raw system logs into an attributed provenance graph that supports subsequent analysis.

\vspace{-9pt}
\subsubsection{Graph Construction}
We parse audit logs to construct a directed multi-graph $G=(V,E)$, 
where nodes represent system entities and edges represent causal or information-flow events among them (e.g., \texttt{read}). 
Each node is associated with a categorical \textit{type}, a set of \textit{semantic attributes (e.g., command line)}, and a \textit{timestamp}, 
while each edge is annotated only with its \textit{type} and \textit{timestamp}. 
Following standard practice~\cite{orthrus,velox}, we focus on three major types of entities and their corresponding interactions: 
\texttt{process}, \texttt{file}, and \texttt{netflow}. 
A complete list of the considered entity and event types is provided in Appendix~\ref{apd:entity_event_types}.

\vspace{-9pt}
\subsubsection{Graph Featurization}
We adopt distinct featurization for nodes and edges to encode type and semantic properties.

\noindent\ding{172}\textbf{Node Features.}
For each node $v$, we construct a feature vector $\mathbf{x}_v$ by concatenating a one-hot vector encoding its type (e.g., \textit{process}) with a semantic attribute embedding derived from textual information such as file paths or command lines.  
Following prior work~\cite{flash, orthrus, zhang2020dynamicmalwareanalysisfeature}, the attribute embedding is obtained using Word2Vec~\cite{word2vec,leimeister2019skipgramwordembeddingshyperbolic}:
\vspace{-5pt}
\[
\mathbf{x}_{v,\text{attr}}=Word2Vec(x_{text}),\quad
\mathbf{x}_v=\big[\text{one-hot(type)}~\|~\mathbf{x}_{v,\text{attr}}\big]
\]
\noindent\ding{173}\textbf{Edge Features.}
For each edge $(u,v)$, the feature vector $\mathbf{e}_{uv}$ is represented as a \textit{multi-hot} indicator over event types (e.g., \texttt{read}).  
This representation captures the fact that multiple event types may occur between the same pair of entities, preserving the diversity of interactions \rev{while substantially reducing edge count; we empirically validate that this aggregation does not degrade detection in Appendix~\ref{sec:multihot_ablation}}.

\vspace{-9pt}
\subsection{Multi-view Anomaly Characterization}
\vspace{-9pt}
\label{sec:multi_view_characterization}
Rather than collapsing all information into a single representation and then obtaining a final anomaly score, 
we compute three per-view anomaly scores for each node $v\in V$: 
$S_{\text{attr}}(v)$, $S_{\text{struc}}(v)$, and $S_{\text{causal}}(v)$, then combine the three scores through a simple voting-based fusion approach (Section~\S\ref{sec:fusion_detection}).
Each view is modeled and scored independently for two main reasons. \textit{(1) Disentanglement and contamination control.}  
In our preliminary experiments, we observed that message passing~\cite{messagepassing} in GNNs can allow attribute-level novelty to spread through the graph, making normal structural patterns appear abnormal.  
This observation originally motivated separating the node-centric perspective into attribute and structural views (see ablation study in \S\ref{sec:detector_enhancement}).  
Per-view scoring keeps each view calibrated to its own feature distribution and helps mitigate cross-signal interference. 
\textit{(2) Task heterogeneity and learning stability.}
The three views differ primarily in their analytical objectives rather than in the input features they use.  
While all views rely on the same provenance graph, each defines a distinct prediction task—attribute outlier detection, structural pattern assessment, and edge-type plausibility estimation.  
A unified optimization objective would force these heterogeneous tasks to share a single loss landscape, 
where the gradients associated with one objective could interfere with those of another. 
Such coupling risks obscuring the semantics of each task and destabilizing the learning process, 
ultimately producing a score whose meaning is difficult to interpret.  
Modeling and scoring each view separately avoids these conflicts and allows each view to focus on its respective objective; 
exploring more integrated optimization strategies is left for future work.

\vspace{-9pt}
\subsubsection{View 1: Node-centric Attribute Anomaly}\ \ \ \ 
\label{sec:view_attribute}

\noindent\textbf{Goal.} The attribute view focuses on detecting anomalies based on an entity’s semantic attributes.

\noindent\textbf{Method.}
This view focuses purely on each entity's semantic attributes, independent of its structural context. 
Given the attribute vector $\mathbf{x}_{v,\text{attr}}$ obtained in Section~\S\ref{sec:graph_preparation}, we measure how far a node’s semantics deviate from those observed in benign data. 
Specifically, we employ a density-based K-Nearest Neighbor (KNN) detector~\cite{knn}, which estimates the local density of a sample by computing its average distance to the $k$ nearest neighbors in the semantic feature space. 
\textit{A node is assigned a higher anomaly score $S_{\text{attr}}(v)$ when its representation lies in a sparse region—indicating that it is semantically dissimilar to most other entities}. 

\noindent\textbf{Rationale.} 
This design avoids assumptions about the underlying data distribution and can be efficiently accelerated on GPUs~\cite{douze2024faiss,johnson2019billion}, providing both efficiency and scalability for large provenance graphs.

\vspace{-9pt}
\subsubsection{View 2: Node-centric Structural Anomaly}\ \ \ \ 
\label{sec:view_structural}

\noindent\textbf{Goal.}
The structural view focuses on detecting anomalies from the perspective of an entity’s role and structure pattern, independent of its semantic attributes.

\noindent\textbf{Method.}
To capture purely structural behavior, we construct a simplified \textbf{type-only provenance graph} $G_{\text{type}}$, where each node is represented as a one-hot vector and each edge is represented as a multi-hot event-type vector defined in Section~\S\ref{sec:graph_preparation}. 
This abstraction removes semantic noise while retaining the topological information needed to learn normal behavior patterns.

We train a self-supervised \textbf{Graph Masked Autoencoder (GMAE)}~\cite{graphmae} on $G_{\text{type}}$ to model typical structures patterns.
Its encoder employs an \textbf{edge-aware graph-attention} mechanism~\cite{gatv2} that integrates event-type information into message passing:
\vspace{-10pt}
\begin{align}
c_{uv} &= \text{attn-MLP}\!\left(\mathbf{h}_u~\|~\mathbf{h}_v~\|~\mathbf{e}_{uv}\right), \label{eq:attn_score}\\
\alpha^{(l)}_{uv} &= 
\frac{\exp(c_{uv})}
     {\sum_{k\in\mathcal{N}_v\cup\{v\}}\exp(c_{vk})}, \label{eq:softmax}\\
\mathbf{h}^{(l+1)}_v &=
\sigma\!\left(
  \sum_{u\in\mathcal{N}_v\cup\{v\}}
     \alpha^{(l)}_{uv}\,
     \mathbf{W}^{(l)}_{\text{val}}\mathbf{h}^{(l)}_u
\right). \label{eq:node_update}
\end{align}
where $\mathcal{N}_v$ represents neighbors of node $v$,  and $h_v^{l}$ represents the representation of node $v$ at layer $l$.
Including edge features allows the encoder to adjust attention weights according to event semantics, improving context awareness without additional supervision.

During training, a subset of nodes is randomly masked, and the decoder reconstructs their original type vectors from the encoder outputs.  
The reconstruction objective is optimized using a \textbf{cosine-similarity loss}, which encourages the reconstructed vectors to align with the ground-truth representations.  
This self-supervised objective promotes learning of common structural relations among entity types.  
After training, the encoder output $\mathbf{h}_{\text{struc}}(v)$ serves as the structural embedding for node $v$. 
We then estimate its anomaly score $S_{\text{struc}}(v)$ using a KNN density estimator~\cite{knn} in the embedding space.  
\textit{Nodes that are structurally dissimilar to most others—i.e., having large distances from their nearest neighbors—are assigned higher anomaly scores.}

\noindent\textbf{Rationale.}
By focusing solely on structural patterns, which are captured through the connectivity between entity types and their associated event types, this view reduces the influence of attribute semantics on structural embeddings and helps prevent benign yet novel attribute variations from causing normal structures to be misidentified as abnormal. 
In this way, it provides a clear basis for detecting deviations from typical structural patterns.

\vspace{-9pt}
\subsubsection{View 3: Edge-centric Causal Anomaly}\ \ \ \ 
\label{sec:view_causal}

\noindent\textbf{Goal.}
The causal view focuses on detecting anomalies in the plausibility of interactions between entities—that is, whether the edge linking two nodes aligns with normal causality.

\noindent\textbf{Method.}
To evaluate interaction plausibility, we formulate an edge-type prediction task on the full, attribute-rich provenance graph $G_{\text{full}}$. 
We reuse the \textbf{Graph Masked Autoencoder (GMAE)}~\cite{graphmae} framework from the structural view but train it on richer semantic features ($\mathbf{x}_v$) with attributes to obtain semantically informed embeddings $\mathbf{h}_{\text{sem}}(v)$. 
After self-supervised pretraining, a lightweight 2-layer MLP~\cite{mlp} decoder predicts the event types for each edge $(u,v)$ using the concatenated embeddings of its involved nodes:
\vspace{-5pt}
\begin{equation}
\hat{\mathbf{y}}_{uv} = 
\text{MLP}\!\left([\mathbf{h}_{\text{sem}}(u)~\|~\mathbf{h}_{\text{sem}}(v)]\right).
\label{eq:edge_prediction}
\end{equation}

Since an edge may correspond to multiple event types, 
the ground truth $\mathbf{y}_{uv}$ is represented as a \textit{multi-hot} vector,  
and the model is trained on benign data using a \textbf{weighted binary cross-entropy (BCE)} loss 
$\mathcal{L}_{\text{BCE}}(\hat{\mathbf{y}}_{uv}, \mathbf{y}_{uv})$.  
The weight for each event type~$i$ is computed from the benign training data as 
$w_i = \sqrt{(N-n_i)/n_i}$,  
where $N$ is the total number of edges and $n_i$ is the count of class~$i$, 
to mitigate class imbalance during training.
During inference, the prediction loss of each edge serves as its anomaly score, and a node’s causal anomaly score $S_{\text{causal}}(v)$ is defined as the maximum loss among its adjacent edges, highlighting entities involved in implausible interactions.  
\textit{Edges (and the associated nodes) that yield higher prediction losses are therefore considered more anomalous.}

\noindent\textbf{Rationale.}
This view models how entities interact through events and evaluates whether such interactions are behaviorally plausible. 
By learning interaction semantics through an event-type prediction objective on benign data, 
it captures behavioral regularities distinct from those characterized in the attribute and structural views. 
This separation thus allows the framework to identify abnormal causal flows—such as unexpected information transfers or execution chains—by detecting deviations from these learned patterns.

\vspace{-9pt}
\subsection{A Multi-Dimensional Anomaly Fusion and Voting-based Detection Framework}
\vspace{-9pt}
\label{sec:fusion_detection}

\noindent\textbf{Goal.}
Given three view-specific anomaly scores per node—attribute ($S_{attr}$), structure ($S_{struc}$), and causality ($S_{causal}$)—our objective is to integrate them into a unified, transparent decision framework. 
The design should (i) normalize heterogeneous score scales, 
(ii) combine anomaly scores from multiple views under a principled rule space, 
and (iii) provide transparency for post-hoc diagnosis. 
All thresholds are derived solely from benign validation data without reliance on test  data~\cite{arp2021dosdontsmachinelearning,learningfromdata}.

\noindent\textbf{Method.}
Our approach consists of three steps: \noindent\ding{172} percentile normalization, \noindent\ding{173} fusion through seven anomaly dimensions, and \noindent\ding{174} ensemble voting.

\noindent\ding{172}\textbf{Step 1: Percentile Normalization.}
For each $\text{view}\in\{\text{attr},\text{struc},\text{causal}\}$, 
we map the raw score of node $v$ to its empirical quantile rank within the benign validation distribution:
\vspace{-5pt}
\begin{equation*}
S'_{\text{view}}(v)=
\text{Quantile}\!\left(S_{\text{view}}(v)\mid S_{\text{view}}^{\text{benign}}\right)\in[0,1].
\end{equation*}
This converts heterogeneous scores into a comparable, scale-free space, 
where a higher $S'_{\text{view}}(v)$ consistently reflects a stronger relative anomaly signal.

\noindent\ding{173}\textbf{Step 2: Seven Anomaly Dimensions.}
We define seven binary detectors $\{D_i\}_{i=1}^{7}$, each corresponding to a specific linear combination pattern of the three normalized scores.
A detector flags node $v$ when its combined score is not smaller than a threshold $\tau_i$ that is set as the\textit{ maximum benign score} for that dimension.  
The detectors are organized by the number of contributing views.

\noindent\textbf{Group 1 - Single-view Specialists (D1–D3):}
capture strong anomalies confined to one view, where
$D1=[S'_{\text{attr}}(v)\geq\tau_1]$, 
$D2=[S'_{\text{struc}}(v)\geq\tau_2]$, and 
$D3=[S'_{\text{causal}}(v)\geq\tau_3]$.

\noindent\textbf{Group 2 – Pairwise Corroboration (D4–D5):}  
capture cases where two views jointly indicate abnormality.  
Let $S'_{\max1}(v)$ and $S'_{\max2}(v)$ denote the two largest values among 
$\{S'_{\text{attr}},S'_{\text{struc}},S'_{\text{causal}}\}$.
% \vspace{-5pt}
\[
D4:~[S'_{\max1}(v)+S'_{\max2}(v)\geq\tau_4].
\]
For a weighted variant, we compute a dynamic fusion with a weighting coefficient $\alpha$, which is set to 5 by default. A hyperparameter study of $\alpha$ is reported in Section~\S\ref{sec:hyperparameter}.
\vspace{-5pt}
\begin{equation*}
\small
w_i=\frac{e^{\alpha S'_i(v)}}{\sum_{j\in\text{Top-2}}e^{\alpha S'_j(v)}},\quad
S_{\text{fused}}=\sum_{i\in\text{Top-2}} w_iS'_i(v),
\end{equation*}
\vspace{-5pt}
\vspace{-5pt}
\[
D5:~[S_{\text{fused}}\geq\tau_5].
\]

\noindent\textbf{Group 3 – All-View Aggregation (D6–D7):}  
represent joint evidence from all three views:
\vspace{-5pt}
\[
D6:~[S'_{\text{attr}}(v)+S'_{\text{struc}}(v)+S'_{\text{causal}}(v)\geq\tau_6],
\]
and its weighted counterpart:
\vspace{-5pt}
\begin{equation*}
\small
w_i=\frac{e^{\alpha S'_i(v)}}{\sum_{j\in\{\text{attr,struc,causal}\}} e^{\alpha S'_j(v)}},\quad
S_{\text{fused}}=\sum_{i} w_iS'_i(v),
\end{equation*}
\vspace{-5pt}
\[
D7:~[S_{\text{fused}} \geq \tau_7].
\]

\noindent\ding{174}\textbf{Step 3: Ensemble Voting.}
Each detector provides a binary decision, and their collective votes determine the final outcome:
\vspace{-5pt}
\[
\small
\begin{aligned}
V(v) &= \sum_{i=1}^{7}[D_i(v)\text{ flags}],\\
\text{is\_malicious}(v) &= \text{True} \iff V(v)\ge T_v.
\end{aligned}
\] 
We use $T_v=4$ by default, corresponding to a simple majority among the seven detectors.
A sensitivity analysis on $T_v$ is reported in the evaluation (see Section~\S\ref{sec:voting_effectiveness}).

\noindent\textit{Analyst-facing output.} 
For each alerted node, we report both the normalized triplet 
$(S'_{\text{attr}},S'_{\text{struc}},S'_{\text{causal}})$ 
and the detector outcomes $[D_1,\dots,D_7]$ with $V(v)$, 
so that analysts can trace which evidence sources and thresholds jointly contributed to the decision.

\noindent\textbf{Rationale.}
To effectively integrate the three heterogeneous anomaly scores, we first normalize them. We adopt percentile normalization as it provides a scale-free and distribution-agnostic basis for integration. Section~\S\ref{sec:norm_ablation} confirms that this choice is critical for a stable, balanced fusion.

Building on this foundation, we adopt a linear-fusion paradigm because it naturally preserves the monotonic relationship among anomaly scores—where higher view-specific scores consistently indicate stronger anomaly evidence—and helps keep the contribution of each view directly transparent. 
In contrast, density- or probability-based integration emphasizes distributional deviations, which may violate this monotonic property and assign high anomaly likelihoods to samples with moderate scores simply because they lie in sparse regions. 
This theoretical rationale guided our design choice, and subsequent evaluation (Section~\S\ref{sec:fusion_rationale}) was consistent with our analysis, showing that linear fusion remained stable and effective across datasets, whereas density- or probability-based alternatives tended to exhibit the expected sensitivity to distributional variations.

Building on this principle, we adopt linear fusion to combine the scores from the three views. 
Conceptually, any linear-fusion strategy can be grouped into three categories according to the number of participating views—using one, two, or all three—which correspond to Groups~1–3 in our design. 
Single-view detectors (Group~1) apply a direct threshold on individual scores, whereas the multi-view detectors in Groups~2 and~3 are implemented in both unweighted and adaptive weighted forms. 
Accordingly, we design seven detectors that form a small family of representative linear-combination patterns among the three normalized scores, capturing single-view, pairwise, and all-view anomaly evidence. 
All detectors follow a monotonic reasoning principle: higher view-specific scores indicate stronger anomaly evidence. 
For instance, Group~2 fuses the two highest-scoring views, and in its adaptive variants larger weights are assigned to stronger signals, reflecting this monotonic rule. 
All detectors remain lightweight; the weighted forms use only a single weighting coefficient~$\alpha$.

Together, these outputs clarify how multi-view evidence is combined and calibrated, 
offering decision-level transparency that helps analysts trace the origin of each alert 
and perform post-hoc diagnosis or auditing when needed.

\vspace{-9pt}
\section{Evaluation}
\vspace{-9pt}
\label{sec:evaluation}

In this section, we evaluate \system{} in comparison with recent state-of-the-art (SOTA) provenance-based intrusion detection systems to assess its detection performance and efficiency. 
The evaluation provides a systematic and transparent analysis of how well \system{} detects attacks and operates under realistic system settings. 
We first introduce the research questions that guide our study and then present the experimental setup, including the datasets, baselines, and evaluation metrics.

We investigate the following four questions:

\noindent\textbf{RQ1: Effectiveness.}  
How effective is \system{} in detecting a broad range of attacks while maintaining comparatively low false positives relative to recent SOTA systems?

\noindent\textbf{RQ2: Component analysis.}  
How do the key design components of \system{}—particularly the multi-view analysis, and the multi-dimensional anomaly fusion and detection framework—shape its overall detection behavior?

\noindent\textbf{RQ3: Efficiency.}  
Is \system{} computationally efficient and scalable?

\noindent\textbf{RQ4: Hyperparameter study.}  
How do key hyperparameters—such as the weighting coefficient $\alpha$, learning rate—affect the overall performance of \system{}?

\vspace{3pt}
\noindent\textbf{Implementation.}
\system{} is implemented in Python. We use Word2Vec~\cite{word2vec} for attribute embedding and adopt a Graph Masked Autoencoder (GMAE)~\cite{graphmae} with edge-aware graph attention~\cite{gatv2} as the graph learning backbone. 
Source code and configurations will be released upon publication to facilitate reproducibility. 
All experiments were conducted with fixed random seeds and consistent data splits following prior work~\cite{velox,orthrus}. 
Key hyperparameters were selected based on validation performance for each dataset, following the same tuning protocol for all compared methods. 
A detailed hyperparameter study is presented in Section~\S\ref{sec:hyperparameter}. 

All experiments were performed on a server running Ubuntu~22.04, equipped with a 128-core AMD EPYC~7543 processor, 512\,GB of RAM, and an NVIDIA A100 GPU.

\vspace{-9pt}
\subsection{Experimental Setup}
% \vspace{-9pt}
\vspace{-9pt}
\subsubsection{Datasets}
Following recent SOTA works~\cite{orthrus, velox}, we use large-scale datasets from the DARPA Transparent Computing (TC) program~\cite{darpa3program}, which serve as standard benchmarks for PIDSs. Other public datasets are either substantially smaller in scale~\cite{streamspot, unicorn, atlas} or not publicly accessible at the time of writing~\cite{provdetector, watson}.

The DARPA TC datasets originate from red-team engagements that simulated realistic, multi-stage attacks (e.g., intrusions on SSH, email, and web servers) amid a high volume of benign user activities. Data were collected across three operating systems—FreeBSD~\cite{freebsd}, Linux~\cite{linux}, and Android~\cite{android}—using distinct capture agents: CADETS, THEIA, and CLEARSCOPE (abbreviated as \textbf{\texttt{CLRSCP}} in tables when space is limited).  \rev{To assess generalization beyond the TC family, we evaluate on the DARPA Operationally Transparent Cyber (OpTC)~\cite{darpaoptc} dataset---an independent program collected via Windows ETW at enterprise scale (500 hosts, 10 days). Following prior work~\cite{velox}, we use the three targeted hosts (\texttt{H201}, \texttt{H501}, \texttt{H051}) as evaluation subsets.} A salient characteristic of these datasets is the severe class imbalance, where malicious events are vastly outnumbered by benign ones. Table~\ref{tab:dataset_statistics} in Appendix~\ref{apd:statistics} provides detailed statistics.

\noindent\textbf{Data Splitting.} We follow the splitting protocols adopted in prior work~\cite{orthrus, velox} to enable comparable evaluation, using consistent train/validation/test partitions as specified by the corresponding benchmarks~\cite{velox}. The details of our data splitting are provided in Appendix~\ref{apd:dataset_splitting}.

\vspace{-9pt}
\subsubsection{Compared Baselines}
\label{sec:compared_baseline}
We compare \system{} with six representative and publicly reproducible PIDSs: Kairos~\cite{kairos}, NodLink~\cite{nodlink}, MAGIC~\cite{magic}, FLASH~\cite{flash}, Orthrus~\cite{orthrus}, and Velox~\cite{velox}.  
We exclude SLOT~\cite{slot} and R-CAID~\cite{rcad} because their official implementations were unavailable during this study; moreover, the reported performance of SLOT on the DARPA datasets is comparable to MAGIC and FLASH, so its exclusion is unlikely to affect conclusions.  
For each baseline, we started from the configurations recommended in the corresponding papers or repositories and further performed validation-based tuning under the same protocol as \system{}. \rev{Other SOTA PIDSs are subgraph-level~\cite{ocr-apt} or rule-based~\cite{captain}, and are therefore excluded.}
We report the best-performing results observed across both sources to ensure fairness and reproducibility. 
 \rev{We additionally include a score-level ensemble baseline, \textbf{Magic+Velox} (M+V in tables), which linearly fuses the per-node anomaly scores of Magic and Velox to test whether \system{}'s gains can be obtained by trivially combining existing detectors.}

\vspace{-9pt}
\subsubsection{Evaluation Metrics}
Given the extreme data imbalance and the operational constraints of a practical IDS, standard metrics like AUC~\cite{auc} and accuracy can be misleading (e.g., a system with 99.99\% accuracy may still produce an impractically high number of false alarms).
We therefore report the Matthews Correlation Coefficient (MCC) as a more informative thresholded metric for imbalanced binary classification. 
Because any thresholded metric ultimately depends on the chosen anomaly threshold, we further complement MCC with a threshold-independent, ranking-oriented measure—Area under the Detection–Precision curve (ADP)—to assess how well a method prioritizes malicious entities above benign ones under varying precision levels.

\noindent\textbf{Matthews Correlation Coefficient (MCC)~\cite{mcc}.}
MCC summarizes binary classification quality and remains robust under severe imbalance by accounting for all four entries of the confusion matrix:
% \vspace{-5pt}
{\small
\[
\text{MCC}=\frac{TP \times TN - FP \times FN}{\sqrt{(TP+FP)(TP+FN)(TN+FP)(TN+FN)}}
\]
}
An MCC of $+1$ indicates perfect prediction, $0$ random prediction, and $-1$ inverse prediction.

\noindent\textbf{Area under the Detection–Precision Curve (ADP).}
Proposed by Velox~\cite{velox}, ADP evaluates how effectively a system detects attacks across a spectrum of precision levels, mitigating reliance on any single threshold. 
Let $D(p)$ denote the fraction of detected attacks at precision $p$. The Detection–Precision curve plots $D(p)$ against $p$, and
\vspace{-5pt}
\[
\text{ADP} = \int_0^1 D(p)\, dp,\quad 
D(p) = \frac{|\{A_i \mid A_i \cap R(p) \neq \emptyset\}|}{k},
\]
where $k$ is the number of attack campaigns, $A_i$ is the set of malicious nodes for attack $i$, and $R(p)$ is the set of nodes flagged as anomalous when the decision threshold is swept to achieve precision $p$.

\textbf{Ranking for \system{}.}
Since \system{} produces seven detector decisions from three normalized view-specific scores, we derive a unified ranking by first ordering nodes according to their vote count $V(v)=\sum_{i=1}^7 D_i(v)$, with ties resolved using the maximum score among the three views. This process yields a deterministic total order for ADP computation.

\newcommand{\solidcircle}{\ding{51}}  % Defines solid circle
\newcommand{\hollowcircle}{\ding{55}}

\vspace{-9pt}
\subsection{Overall Detection Performance (RQ1)}
\vspace{-9pt}
\label{sec:rq1}

To answer our first research question, we evaluate \system{}’s detection performance under the refined ground truth introduced below, focusing on two aspects: (1) its ability to detect every attack campaign (attack coverage), and (2) its precision in identifying individual malicious entities (node-level performance).

\vspace{-5pt}
\subsubsection{Refined Ground Truth Construction}
\vspace{-5pt}
\label{sec:refinedgt}

While Orthrus~\cite{orthrus} provides the most comprehensive ground truth available for the DARPA TC datasets, our cross-check with the official DARPA engagement reports~\cite{Ground-truth_file} and the REAPr dataset~\cite{newlabel} revealed that several documented campaigns (e.g., \textit{Phishing Email w/ Executable Attachment} in THEIA-E3) were not represented in the released labels. 
This observation indicates that even strong benchmarks still leave room for refinement.

The refinement was conducted \emph{before} any model training and is independent of our method design. 
Guided by the DARPA reports and REAPr, we manually examined the corresponding audit logs to identify and verify missing entities using multiple evidence sources—timestamp alignment, filename or command-line matching, file-path consistency, and causal dependencies—with cross-review among annotators. 
This process required extensive manual inspection and verification effort, and it only supplements verifiable malicious entities without altering any original Orthrus labels.

In total, 21 additional malicious nodes were confirmed across three datasets—THEIA-E3 (+11), CLEARSCOPE-E3 (+8), and CLEARSCOPE-E5 (+2). 
We use this refined ground truth as the primary reference for evaluation, as it more comprehensively reflects all attack activities documented in the official engagements. 
For completeness, results under the original Orthrus labels are also provided in Appendix~\ref{apd:orthrus_results}. 
The newly added malicious nodes (including UUID, attributes, and timestamps) are listed in Appendix~\ref{apd:new_ground_truth} for transparency and will be publicly released to support community benchmarking and future research on PIDSs.

\vspace{-5pt}
\subsubsection{Attack Campaign Coverage}
\vspace{-5pt}

An effective detection system should detect every attack campaign without omission. 
We consider an attack ``detected’’ if at least one node directly involved in the campaign is flagged as malicious. 
Table~\ref{tab:attack_coverage} summarizes the results. 
\system{} successfully detects all attacks across \rev{nine} datasets except for a single campaign in CLEARSCOPE-E5, achieving near-complete coverage across diverse environments. This result highlights \system{}’s ability to capture diverse attack campaigns.

\noindent\textbf{Analysis of the missed campaign.}
The missed case, ``Lockwatch APK Java APT,’’ was exceptionally stealthy. 
Inspection of its anomaly scores shows that the involved nodes were already ranked near the decision boundary. 
For instance, one node yielded ($S_{\text{attr}}=0.401$, $S_{\text{struc}}=0.994$, $S_{\text{causal}}=0.390$), suggesting that the attack effectively mimicked benign attributes and causal flows while exhibiting a distinct structural deviation. 
Another node in the same campaign had ($S_{\text{attr}}=0.983$, $S_{\text{struc}}=0.984$, $S_{\text{causal}}=0.988$), with its aggregated score slightly below the detection threshold. 
These results indicate that \system{} was responsive to this campaign’s abnormal structure, and a slightly relaxed threshold could have captured it.
This single miss reflects a conservative thresholding choice made to reduce FPs.

\begin{table}
  \centering
  \begin{small}
  \begin{tblr}{
    rowsep = 0pt,
    colsep = 0.5pt,
    cells = {c},
    width = \linewidth,
    colspec = {Q[150]Q[44]Q[44]Q[44]Q[44]Q[44]Q[44]Q[44]Q[44]Q[44]Q[44]Q[44]Q[44]Q[44]Q[44]Q[44]Q[44]Q[44]},
    cell{1}{1}  = {r=2}{},
    cell{1}{2}  = {c=5}{},
    cell{1}{7}  = {c=4}{},
    cell{1}{11} = {c=5}{},
    cell{1}{16} = {r=2,c=3}{},
    cell{2}{2}  = {c=3}{},
    cell{2}{5}  = {c=2}{},
    cell{2}{7}  = {c=3}{},
    cell{2}{12} = {c=4}{},
    % cell{3}{16} = {fg=diffblue},
    % cell{3}{17} = {fg=diffblue},
    % cell{3}{18} = {fg=diffblue},
    % cell{4-11}{16} = {fg=diffblue},
    % cell{4-11}{17} = {fg=diffblue},
    % cell{4-11}{18} = {fg=diffblue},
    % row{10} = {fg=diffblue},
    vline{2}            = {-}{},
    vline{5,7,10,11,12} = {-}{},
    vline{16}           = {-}{fg=diffblue},
    hline{1,12}         = {-}{0.1em},
    hline{2,3,4}        = {-}{},
  }
  Dataset$\rightarrow$ & CADETS &    &    &    &    & THEIA &    &    &    & CLEARSCOPE &    &    &    &    & OpTC &      &      \\
                       & E3     &    &    & E5 &    & E3    &    &    & E5 & E3         & E5 &    &    &    &      &      &      \\
  PIDS$\downarrow$     & A1 & A2 & A3 & A1 & A2 & A1 & A2 & A3 & A1 & A1 & A1 & A2 & A3 & A4 & A1 & A2 & A3 \\
  Kairos     & \hollowcircle & \solidcircle  & \hollowcircle & \hollowcircle & \hollowcircle & \solidcircle  & \solidcircle  & \hollowcircle & \hollowcircle & \solidcircle  & \solidcircle
    & \hollowcircle & \hollowcircle & \hollowcircle & \solidcircle & \solidcircle & \solidcircle \\
  Magic      & \solidcircle  & \solidcircle  & \solidcircle  & \solidcircle  & \solidcircle  & \solidcircle  & \solidcircle  & \hollowcircle & \solidcircle  & \solidcircle  & \solidcircle
    & \hollowcircle & \solidcircle  & \solidcircle  & \solidcircle & \solidcircle & \solidcircle \\
  NodLink    & \solidcircle  & \solidcircle  & \solidcircle  & \solidcircle  & \solidcircle  & \solidcircle  & \solidcircle  & \hollowcircle & \solidcircle  & \solidcircle  & \solidcircle
    & \hollowcircle & \solidcircle  & \hollowcircle & \solidcircle & \solidcircle & \solidcircle \\
  Flash      & \hollowcircle & \solidcircle  & \solidcircle  & \solidcircle  & \solidcircle  & \solidcircle  & \solidcircle  & \hollowcircle & \solidcircle  & \solidcircle  & \solidcircle
    & \solidcircle  & \solidcircle  & \solidcircle  & \solidcircle & \solidcircle & \solidcircle \\
  Orthrus    & \hollowcircle & \solidcircle  & \solidcircle  & \hollowcircle & \solidcircle  & \solidcircle  & \hollowcircle & \solidcircle  & \solidcircle  & \solidcircle  & \solidcircle
    & \hollowcircle & \hollowcircle & \hollowcircle & \solidcircle & \solidcircle & \solidcircle \\
  Velox      & \solidcircle  & \solidcircle  & \solidcircle  & \hollowcircle & \hollowcircle & \solidcircle  & \solidcircle  & \solidcircle  & \solidcircle  & \solidcircle  & \solidcircle
    & \hollowcircle & \hollowcircle & \solidcircle  & \solidcircle & \solidcircle & \solidcircle \\
     M+V & \solidcircle  & \solidcircle  & \solidcircle  & \solidcircle  & \solidcircle  & \solidcircle  & \solidcircle  & \solidcircle  & \solidcircle  & \solidcircle  & \solidcircle & \hollowcircle  & \hollowcircle & \solidcircle  & \solidcircle & \solidcircle & \solidcircle  \\
  {\scriptsize ProvFusion} & \solidcircle  & \solidcircle  & \solidcircle  & \solidcircle  & \solidcircle  & \solidcircle  & \solidcircle  & \solidcircle  & \solidcircle  & \solidcircle  & \solidcircle
    & \solidcircle  & \hollowcircle & \solidcircle  & \solidcircle & \solidcircle & \solidcircle
  \end{tblr}
  \vspace{-5pt}
  \caption{Attack coverage of \system{} and state-of-the-art baselines.
  \solidcircle~denotes a detected campaign (at least one malicious node flagged),
  while \hollowcircle~denotes a miss. \system{} successfully detects \rev{16 out of 17} campaigns. We provide the details of each attack in
  Appendix~\ref{apd:detailed_name_and_timea_of_attack}.}
  \label{tab:attack_coverage}
  \vspace{-10pt}
  \end{small}
  \end{table}

\begin{table*}[h]
\centering
\begin{small}
\begin{tblr}{
  rowsep = -0.4pt,
  colsep = 2.4pt,
  cells = {c},
  width = 1.0\linewidth,
  % colspec = {Q[71]Q[54]Q[48]Q[56]Q[58]Q[48]Q[50]Q[54]Q[56]Q[54]Q[48]Q[56]Q[58]Q[48]Q[50]Q[54]Q[56]},
  cells = {c},
  cell{2}{2} = {r=7}{},
  cell{2}{10} = {r=7}{},
  cell{10}{2} = {r=7}{},
  cell{10}{10} = {r=7}{},
  cell{18}{2} = {r=7}{},
  cell{18}{10} = {r=7}{},
  vline{2-3,10-11} = {-}{},
  vline{2,10} = {-}{},
  hline{1} = {-}{0.1em},
  hline{2,10,18} = {-}{},
  hline{26} = {-}{0.1em},
}
System     & Dataset        & TP↑         & FP↓        & TN↑             & FN↓         & \rev{F1}↑          & ADP↑          & MCC↑          & Dataset        & TP↑         & FP↓        & TN↑             & FN↓         & \rev{F1}↑          & ADP↑          & MCC↑          \\
Kairos     & {CADETS \\ E3} & 1           & 959        & 280.6K          & 67          & 0.00          & 0.00          & 0.00          & {CADETS \\ E5} & 0           & 6          & 3.14M           & 123         & 0             & 0.01          & 0             \\
Magic      &                & 22          & 16.5K      & 265.0K          & 46          & 0.00          & 0.01          & 0.02          &                & 28          & 245.3K     & 2.89M           & 95          & 0.00          & 0.17          & 0.00          \\
NodLink    &                & 18          & 34.3K      & 247.2K          & 50          & 0.00          & 0.49          & 0.01          &                & \textbf{73}          & 755.9K     & 2.38M           & \textbf{50} & 0.00          & 0.03          & 0.01          \\
Flash      &                & 3           & 4.5K       & 277.0K          & 65          & 0.00          & 0.04          & 0.01          &                & 6           & 34.9K      & 3.10M           & 117         & 0.00          & 0.02          & 0.01          \\
Orthrus    &                & 7           & \textbf{1} & 281.5K          & 61          & 0.18          & 0.81          & 0.30          &                & 1           & 8          & 3.14M           & 122         & 0.00          & 0.34          & 0.03          \\
Velox      &                & 9           & \textbf{1} & 281.5K          & 59          & 0.23          & 0.97          & 0.35          &                & 0           & \textbf{2} & 3.14M           & 123         & 0.00          & 0.01          & 0.00          \\
{M+V} 
           &                & {17} & {173} & {281.3K} & {51} & {0.13} & {0.92} & {0.15}
           &                & {7} & {810} & {3.14M} & {116} & {0.02} & {0.09} & {0.02} \\

ProvFusion &                & \textbf{24} & \textbf{1} & \textbf{281.5K} & \textbf{44} & \textbf{0.52} & \textbf{1.00} & \textbf{0.58} &                & 7  & 9          & \textbf{3.14M}  & 116         & \textbf{0.10} & \textbf{0.92} & \textbf{0.16} \\
Kairos     & {THEIA \\ E3}  & 2           & 21         & 701.5K          & 127         & 0.03          & 0.15          & 0.04          & {THEIA \\ E5}  & 0           & 7          & 1.86M           & 69          & 0.00          & 0.00          & 0.00          \\
Magic      &                & 20          & 97.3K      & 604.2K          & 109         & 0.00          & 0.00          & 0.00          &                & \textbf{61}          & 737.3K     & 1.12M           & \textbf{8}  & 0.00          & 0.00          & 0.00          \\
NodLink    &                & 26          & 258.3K     & 443.2K          & 103         & 0.00          & 0.01          & 0.00          &                & 20          & 175.3K     & 1.68M           & 49          & 0.00          & 0.00          & 0.00          \\
Flash      &                & 21          & 251.2K     & 450.3K          & 108         & 0.00          & 0.01          & 0.00          &                & 41          & 316.2K     & 1.54M           & 28          & 0.00          & 0.01          & 0.01          \\
Orthrus    &                & 4           & \textbf{0} & 701.5K          & 125         & 0.06          & 0.67          & 0.18          &                & 1           & 31         & 1.86M           & 68          & 0.02          & 0.30          & 0.02          \\
Velox      &                & 21          & 117        & 701.4K          & 108         & 0.16          & 1.00          & 0.16          &                & 2           & 63         & 1.86M           & 67          & 0.03          & 0.33          & 0.03          \\
{M+V}
           &                & {16} & {44} & {701.4K} & {113} & {0.17} & {0.49} & {0.18}
           &                & {2} & {306} & {1.86M} & {67} & {0.01} & {0.1} & {0.01} \\

ProvFusion &                & \textbf{91} & 2          & \textbf{701.5K} & \textbf{38} & \textbf{0.82} & \textbf{1.00} & \textbf{0.83} &                & 11 & \textbf{2} & \textbf{1.86M}  & 58          & \textbf{0.27} & \textbf{1.00} & \textbf{0.37} \\
Kairos     & {CLRSCP \\ E3} & 14          & 8.5K       & 102.9K          & 35          & 0.00          & 0.01          & 0.02          & {CLRSCP \\ E5} & 1           & 1          & 150.9K          & 52          & 0.04          & 0.33          & 0.10          \\
Magic      &                & \textbf{42}          & 8.4K       & 102.9K          & \textbf{7}  & 0.01          & 0.02          & 0.06          &                & 7           & 8.3K       & 142.7K          & 46          & 0.00          & 0.00          & 0.01          \\
NodLink    &                & 39          & 22.8K      & 88.5K           & 10          & 0.00          & 0.06          & 0.03          &                & 3           & 27.0K      & 123.9K          & 50          & 0.00          & 0.00          & 0.00          \\
Flash      &                & 36          & 11.1K      & 100.2K          & 13          & 0.01          & 0.01          & 0.04          &                & 0           & \textbf{0} & 150.9K          & 53          & 0.00          & 0.00          & 0.00          \\
Orthrus    &                & 2           & \textbf{5} & 111.3K          & 47          & 0.07          & 0.40          & 0.11          &                & 2           & 8          & 150.9K          & 51          & 0.06          & 0.17          & 0.09          \\
Velox      &                & 3           & 623        & 110.7K          & 46          & 0.01          & 0.50          & 0.02          &                & 8           & 13         & 150.9K          & 45          & 0.22          & 0.45          & 0.24          \\
{M+V}
           &                & {1} & {150} & {112.2K} & {48} & {0.01} & {0.16} & {0.01}
           &                & {6} & {58} & {150.9K} & {47} & {0.10} & {0.17} & {0.10} \\

ProvFusion &                & 6  & 7          & \textbf{111.3K} & 43          & \textbf{0.19} & \textbf{0.86} & \textbf{0.24} &                & \textbf{10} & 16         & \textbf{150.9K} & \textbf{43} & \textbf{0.25} & \textbf{0.77} & \textbf{0.27} 
\end{tblr}

    \vspace{-4pt}
\caption{Node-level detection performance on the refined ground truth. \system{} significantly outperforms all baselines in terms of F-1, ADP, and MCC, while maintaining a much lower number of false positives (FP). Best results are in \textbf{bold}.}
    \vspace{-8pt}

\label{tab:main_results}
\end{small}
\end{table*}

\vspace{-5pt}
\subsubsection{Node-level Performance on the Refined Ground Truth}
\label{sec:node_level_performance}
\vspace{-5pt}

We compare \system{} with six provenance-based intrusion detection systems across all six \rev{TC} datasets. 
As summarized in Table~\ref{tab:main_results}, \system{} consistently achieves higher detection performance than all baselines, identifying more true positives while generating fewer false positives—24.8~TPs and 6.2~FPs on average, compared with 7.2~TPs and 136.5~FPs for the next-best system \rev{(i.e., Velox, the strongest baseline by ADP and also competitive in TP/FP)}. This result demonstrates \system{}’s high precision in identifying individual malicious entities. 

\rev{To test whether trivial detector combinations suffice, Table~\ref{tab:main_results} also reports a score-level fusion of MAGIC and Velox; it inherits MAGIC's FP cost without matching \system{}'s TPs, confirming the value of multi-view fusion over naive score combination.}

Many of the newly verified malicious nodes in the refined ground truth were successfully detected not only by \system{} but also by some strong baselines (e.g., Velox or Orthrus), which further supports the validity of these entities and the quality of the refined annotations.

We further verified that the same performance trends hold under the original Orthrus ground truth (Appendix~\ref{apd:orthrus_results}), with only minor differences in absolute values due to the previously missing attacks. 
A qualitative example in Appendix~\ref{apd:case_study} illustrates how \system{}’s alerts align with the recorded attack sequences, helping analysts reconstruct intrusion chains and better interpret attack progression.

\rev{\noindent\textbf{Generalization to OpTC.}
  Table~\ref{tab:optc_results} shows that \system{}'s advantage transfers
  to the DARPA OpTC dataset: node-centric baselines accumulate
  $10^4$--$10^5$ FPs while edge-centric baselines collapse to $< 3$
  TPs, whereas \system{} sustains a usable detection profile (average of 7 TPs
  / 13 FPs across the three hosts), confirming that multi-view fusion
  remains effective under different logging infrastructure and attack
  distributions.} \rev{Furthermore, a benign-shift stability study (Appendix~\ref{apd:benign_shift}) shows that rotating the validation day has minimal impact on detection.}

\rev{\noindent\textbf{FP/FN Analysis.}
To illustrate how multi-view fusion resolves node/edge-specific FP/FN patterns, we compare \system{} with MAGIC and Velox, two SOTA node- and edge-centric baselines. The two baselines exhibit complementary false negatives: MAGIC misses \texttt{/tmp/mozilla\_admin0/jAG\_iSHt.bin.part}, whose subgraph resembles common temporary-file activity, whereas Velox detects it from its anomalous edge context. Conversely, Velox misses the network endpoint \texttt{128.55.12.110:44354$\rightarrow$8:0}, whose surrounding interactions appear syntactically ordinary, whereas MAGIC flags it due to its atypical structural footprint. \system{} recovers both through cross-view voting. False positives, however, do not accumulate under fusion: a node that appears anomalous in one view is often supported as benign by the others, causing its combined evidence to remain below the voting threshold. True attack nodes typically lack such conflicting benign evidence; even when no single view is decisive, all views assign at least moderate suspicion, allowing consensus to emerge. This difference explains why \system{} improves recall while suppressing false positives, rather than trading one for the other.}
  
\vspace{-5pt}
\begin{tcolorbox}[colback=gray!5!white, colframe=gray!70!black, boxrule=0.7pt, arc=3pt, width=\linewidth]
\textbf{RQ1 Answer:} 
Across six datasets, \system{} achieves comprehensive attack coverage and strong node-level precision while maintaining a low false-positive rate. 
The refined ground truth—constructed independently from method design and validated through official DARPA records—provides a more complete benchmark for future research. 
Overall, the results confirm that \system{} effectively detects diverse and stealthy attacks in provenance graphs under realistic conditions.
\end{tcolorbox}
\vspace{-5pt}

% =====================================================================
% Table 3 — OpTC results (Concern 1, response to "exclusively DARPA TC")
% Place immediately after Table 2 in §5.2.
% =====================================================================
\rev{
\begin{table}[t]
\centering
\small
\setlength{\tabcolsep}{4pt}
\setlength{\aboverulesep}{0pt}
\setlength{\belowrulesep}{0pt}
\renewcommand{\arraystretch}{0.92}
  \begin{tabular}{>{}l>{}l>{}r>{}r>{}r>{}r>{}r}
\toprule
Dataset & Method & TP$\uparrow$ & FP$\downarrow$ & \rev{F1}$\uparrow$ & ADP$\uparrow$ & MCC$\uparrow$ \\
\midrule
\multirow{8}{*}{\shortstack[l]{H051}}
 & Kairos              &    1 &      \textbf{3} & 0.02 & 0.50 & 0.05 \\
 & MAGIC               &  104 & 106K & 0.00 & 0.20 & 0.06 \\
 & NodLink             &   36 &  96.4K & 0.00 & 0.06 & 0.02 \\
 & FLASH               &   27 & 513K & 0.00 & 0.14 & 0.00 \\
 & Orthrus             &    1 &   8 & 0.02 & 0.33 & 0.03 \\
 & Velox               &   2 &   4 & 0.03 & 0.43 & 0.08 \\
  & M+V                 & 5 & 254 & 0.03 & 0.05 & 0.03 \\
 & \textsc{ProvFusion} & \textbf{16} & 30 & \textbf{0.20} & \textbf{0.94} & \textbf{0.22} \\
\midrule
\multirow{8}{*}{\shortstack[l]{H201}}
 & Kairos              &    2 &   \textbf{3} & 0.00 & 1.00 & 0.02 \\
 & MAGIC               & \textbf{2.48K} & 594K & 0.01 & 0.33 & 0.65 \\
 & NodLink             & 1.40K &  77.7K & \textbf{0.03} & 0.07 & \textbf{0.76} \\
 & FLASH               &   751 & 168K & 0.01 & 0.45 & 0.21 \\
 & Orthrus             &    1 &    7 & 0.0 & 0.50 & 0.01 \\
 & Velox               &    1 &   9 & 0.0 & 0.33 & 0.01 \\
  & M+V                 & 10 & 1.17K & 0.01 & 0.89 & 0.10 \\
 & \textsc{ProvFusion} & 3 & 5 & 0.00 & \textbf{1.000} & 0.02 \\
\midrule
\multirow{8}{*}{\shortstack[l]{H501}}
 & Kairos              &    1 &       5 & 0.00 & 0.20 & 0.01 \\
 & MAGIC               &  \textbf{433} & 510K & 0.00 & 1.00 & \textbf{0.14} \\
 & NodLink             &  365 & 102K & \textbf{0.07} & 0.14 & 0.04     \\
 & FLASH               &   31 &  26.6K & 0.00 & 0.01 & 0.01 \\
 & Orthrus             &    1 &    4 & 0.00 & 0.25 & 0.02 \\
 & Velox               &    1 &  7 & 0.00 & 0.33 & 0.01 \\
  & M+V                 & 3 & 7.93K & 0.00 & 0.03 & 0.00 \\
 & \textsc{ProvFusion} & 2 & \textbf{4} & 0.01 & \textbf{1.00} & 0.03 \\
\bottomrule
\end{tabular}
\vspace{-5pt}
 % \captionsetup{labelfont={color=diffblue}, textfont={color=diffblue}}
\caption{Detection results on the DARPA OpTC dataset.}
\label{tab:optc_results}
\vspace{-13pt}
\end{table}
}

\begin{figure*}[t]
  \centering
  \begin{minipage}{0.32\linewidth}
    \centering
    \includegraphics[width=\linewidth]{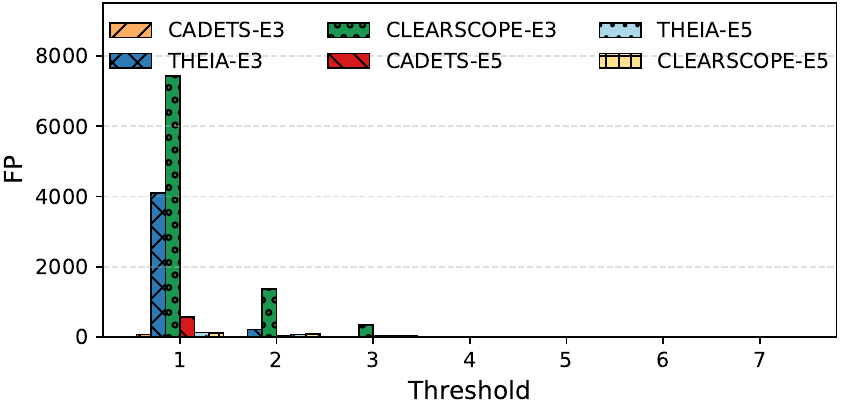}
  \vskip -0.05in
  \vspace{-3pt}
    \caption*{(a) False Positive Count}
        \vspace{-3pt}

    \label{fig:fp}
  \end{minipage}
  \hfill
  \begin{minipage}{0.32\linewidth}
    \centering
    \includegraphics[width=\linewidth]{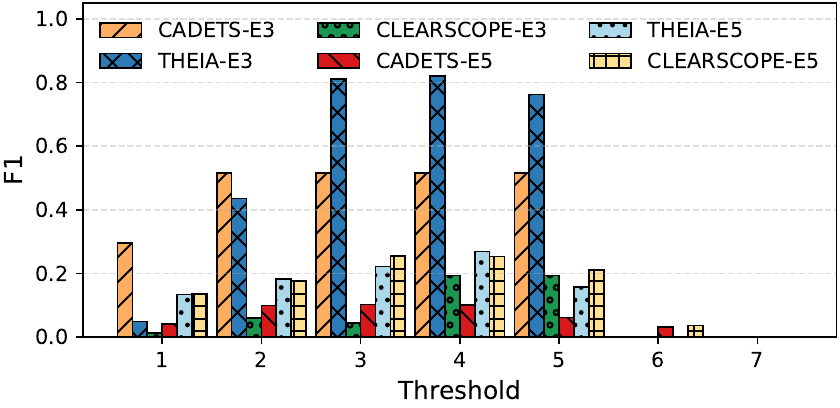}
  \vskip -0.05in
  \vspace{-3pt}
    \caption*{(b) F1 Score}
        \vspace{-3pt}

    \label{fig:f1}
  \end{minipage}
  \hfill
  \begin{minipage}{0.32\linewidth}
    \centering
    \includegraphics[width=\linewidth]{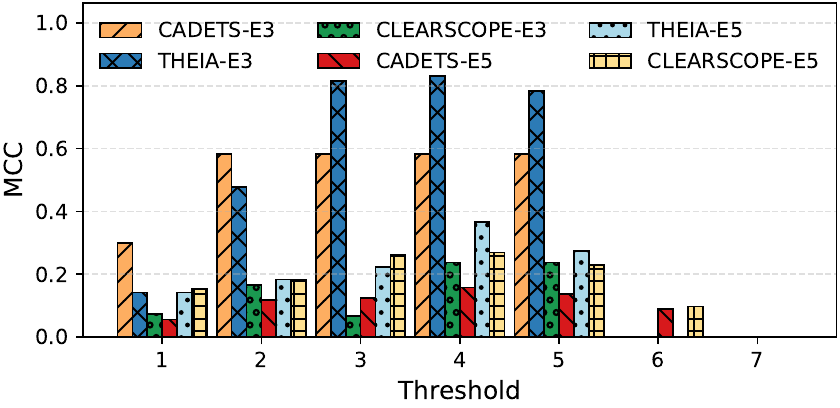}
  \vskip -0.05in
  \vspace{-3pt}
    \caption*{(c) MCC}
        \vspace{-3pt}

    \label{fig:mcc}
  \end{minipage}
  \vskip -0.05in
  \caption{Impact of the voting threshold ($T_v$) on performance. A threshold of $T_v=4$ provides the optimal balance, maximizing the (b) F1 and (c) MCC scores while dramatically reducing the (a) False Positives seen at lower thresholds.}
  \label{fig:threshold_sensitivity}
  \vskip -0.05in
  \vspace{-5pt}

\end{figure*}

\vspace{-9pt}
\subsection{In-depth Analysis (RQ2)}
\vspace{-9pt}
\label{sec:in_depth_analysis}

Having established \system{}’s overall detection performance in Section~\S\ref{sec:rq1}, we now conduct ablation studies to understand the roles of its core design components and the mechanisms driving its effectiveness. 
This section addresses our RQ2: \textit{How do \system{}’s core design choices contribute to its detection capability?}

% --- 这是新增加的章节 ---
\vspace{-5pt}
\subsubsection{Validation of Normalization Strategy}
\vspace{-5pt}
\label{sec:norm_ablation}
We first validate our \textit{a priori} choice of percentile normalization (discussed in \S\ref{sec:fusion_detection}). We compared its performance within the full \system{} framework against three common alternatives: min-max, z-score~\cite{han2011data}, and robust scaling~\cite{robustscaling}. The detailed results are in Appendix~\ref{apd:norm_ablation}.

The evaluation confirms our choice rationale: percentile normalization consistently outperforms all alternatives. \textbf{Min-max} proved highly sensitive to outliers in the validation data, incorrectly compressing the score range of some views (e.g., scaling a 98th-percentile anomaly to only 0.2). This \textbf{silenced their contribution}, leading to missed detections (8/14). \textbf{Z-score} and \textbf{Robust scaling} failed on the heterogeneous, non-Gaussian distributions, as their underlying \textbf{extreme numerical expansion} (e.g., normalized scores $>$ 1000) led to different instabilities. For \textbf{Z-Score}, this expansion forced an inflated detection threshold, which \textbf{silenced true attack signals} and caused a catastrophic loss of detection (5/14). \textbf{Robust scaling} was the most erratic: its expansion mechanism resulted in a dual failure, allowing some views to \textbf{improperly dominate} the fusion (144 FPs) while simultaneously silencing others (1/14 detections).

\begin{table}[t]
\centering
\begin{small}
\begin{tblr}{
  rowsep = 0pt,
  colsep = 0.6pt,
  width = \linewidth,
  colspec = {Q[135]Q[155]Q[84]Q[84]Q[84]Q[84]Q[84]Q[84]Q[108]},
  cells = {c},
  cell{1}{1} = {r=2}{},
  cell{1}{2} = {r=2}{},
  cell{1}{3} = {c=3}{},
  cell{1}{6} = {c=3}{},
  cell{1}{9} = {r=2}{},
  cell{3}{1} = {c=9}{},
  cell{4}{1} = {r=3}{},
  cell{7}{1} = {r=3}{},
  cell{10}{1} = {r=3}{},
  cell{13}{1} = {r=3}{},
  cell{16}{1} = {r=3}{},
  cell{19}{1} = {r=3}{},
  cell{22}{1} = {r=3}{},
  cell{25}{1} = {r=3}{},
  vline{2,3,6,9} = {1,2}{0.05em},
  vline{2-3,6,9} = {4,7,10,13,16,19,22,25}{0.05em},
  vline{2,3,6,9} = {5-6,8-9,11-12,14-15,17-18,20-21,23-24,26-27}{0.05em},
  hline{1,28} = {-}{0.1em},
  hline{3-4} = {-}{0.05em},
  hline{7,10,13,16,19,22,25} = {1}{0.05em},
  hline{7,10,13,16,19,22,25} = {2-9}{0.05em},
}
\textbf{Method} & \textbf{Dataset} & \textbf{MCC $\uparrow$} & & & \textbf{ADP $\uparrow$} & & & \textbf{Detect} \\
 & & E3 & E5 & Avg & E3 & E5 & Avg & Attacks \\
\textit{Individual Model Families (Best over 4 normalizations)} & & & & & & & & \\
AE & CADETS & .571 & .100 & .336 & \textbf{1.00} & .917 & .958 & 5/5 \\
   & CLRSCP & .117 & .291 & .204 & \textbf{1.00} & .553 & .776 & 3/5 \\
   & THEIA  & .735 & \uline{.295} & .515 & \textbf{1.00} & \textbf{1.00} & \textbf{1.00} & 4/4 \\
GMM & CADETS & .396 & \uline{.136} & .266 & .000 & .000 & .000 & 5/5 \\
    & CLRSCP & .150 & .284 & .217 & .000 & .007 & .004 & 4/5 \\
    & THEIA  & .475 & .273 & .374 & .000 & .000 & .000 & 3/4 \\
{Isolation\\Forest} & CADETS & .243 & .101 & .172 & \textbf{1.00} & .917 & .958 & 5/5 \\
                    & CLRSCP & .076 & .189 & .132 & .184 & .310 & .247 & 4/5 \\
                    & THEIA  & .215 & .000 & .108 & \uline{.670} & \uline{.005} & \uline{.338} & 2/4 \\
KNN & CADETS & .523 & .114 & .319 & \textbf{1.00} & .750 & .875 & 4/5 \\
    & CLRSCP & .116 & \textbf{.330} & .223 & .033 & .646 & .339 & 4/5 \\
    & THEIA  & .481 & \uline{.295} & .388 & \textbf{1.00} & \textbf{1.00} & \textbf{1.00} & 4/4 \\
MVG & CADETS & .184 & \uline{.136} & .160 & .720 & \uline{.950} & .835 & 5/5 \\
    & CLRSCP & .157 & .284 & .220 & .333 & .612 & .473 & 4/5 \\
    & THEIA  & .496 & .273 & .384 & \textbf{1.00} & \textbf{1.00} & \textbf{1.00} & 4/4 \\
OCSVM & CADETS & \uline{.582} & .116 & .349 & .000 & .015 & .008 & 5/5 \\
      & CLRSCP & \uline{.247} & \uline{.320} & \textbf{.284} & .000 & .073 & .036 & 4/5 \\
      & THEIA  & .319 & \textbf{.367} & .343 & .000 & .001 & .000 & 4/4 \\
\textbf{Linear} & CADETS & \textbf{.594} & .122 & \uline{.358} & \textbf{1.00} & \textbf{1.00} & \textbf{1.00} & 5/5 \\
                & CLRSCP & \textbf{.303} & .261 & \uline{.282} & \textbf{1.00} & \textbf{1.00} & \textbf{1.00} & 5/5 \\
                & THEIA  & \textbf{.845} & .289 & \uline{.567} & \textbf{1.00} & \textbf{1.00} & \textbf{1.00} & 4/4 \\
\textbf{Voting (Ours)} & CADETS & \uline{.582} & \textbf{.158} & \textbf{.370} & \textbf{1.00} & .925 & \uline{.962} & 5/5 \\
                       & CLRSCP & .238 & .269 & .254 & \uline{.857} & \uline{.774} & \uline{.815} & 4/5 \\
                       & THEIA  & \uline{.831} & \textbf{.367} & \textbf{.599} & \textbf{1.00} & \textbf{1.00} & \textbf{1.00} & 4/4 \\
\end{tblr}
\end{small}
    \vspace{-5pt}
\caption{Best performance of different fusion/detection paradigms vs. our voting mechanism. 
\textbf{Bold}: best per dataset; \uline{underline}: second best. 
All results use the best normalization found in hyperparameter search.}
    \vspace{-10pt}
\label{tab:scorer_comparison}
\end{table}

\vspace{-5pt}
\subsubsection{Ablation: Evaluating the Rationale of the Multi-Dimensional Anomaly Fusion and Detection Framework}
\vspace{-5pt}
\label{sec:fusion_rationale}

As described in \S\ref{sec:fusion_detection}, our fusion and detection framework integrates multi-view anomaly evidence through two stages: 
(i) aggregating view-specific scores using monotonic linear detectors, and 
(ii) combining their outputs through a voting rule.  
This ablation examines whether such a structured score–decision integration contributes to stable and reliable detection across datasets. 
We replace the proposed framework with representative alternatives from major anomaly detection paradigms and compare their performance under the same experimental setting.

\noindent\textbf{Comparative setup.}
We evaluate five model families:  
(1) \textit{linear combination models} (weighted summations of the three scores under different weighting schemes);  
(2) \textit{probabilistic models} (Gaussian Mixture Models~\cite{gmm} and multivariate Gaussian density estimation~\cite{mvg});  
(3) \textit{distance/density-based models} (K-Nearest Neighbor~\cite{knn});  
(4) \textit{boundary-based models} (One-Class SVM~\cite{oneclasssvm} and Isolation Forest~\cite{IsolationForest}); and  
(5) \textit{reconstruction-based deep models} (a lightweight autoencoder using reconstruction error~\cite{autoencoderanomaly}).  
Each family was explored over a broad parameter space (\textbf{including all four normalization methods from \S\ref{sec:norm_ablation}}) and we report its \emph{best observed} result per dataset (Table~\ref{tab:scorer_comparison}).
For instance, the linear combination models enumerated all integer weight triplets $(w_{\text{attr}}, w_{\text{struc}}, w_{\text{causal}})\in\{0,\ldots,19\}^3\setminus\{(0,0,0)\}$—a total of $7{,}999$ combinations. 
Other families varied core hyperparameters such as neighborhood sizes and learning rates (see details in Appendix~\ref{apd:parameterexplore}).  
All models used identical normalized inputs and data partitions for fairness.

\noindent\textbf{Finding 1: linear fusion best supports the monotonicity principle.}
Across all datasets, the \textbf{linear combination models} achieved the most stable and overall highest performance, whereas other paradigms performed well only on certain datasets.  
Methods relying on statistical rarity often assigned high anomaly scores to rare but benign nodes while overlooking nodes whose three scores were all high.  
For example, under Isolation Forest with Percentile normalization, a node with a skewed score vector (e.g., $[0.98,0.50,0.11]$) can receive a higher anomaly score than a node with uniformly high scores (e.g., $[0.99,0.99,0.98]$).
This behavior suggests that when individual scores already convey strong anomaly evidence, monotonic linear fusion provides a more natural and consistent integration.

\noindent\textbf{Finding 2: stability of the proposed fusion across datasets.}
Although linear fusion performed best overall, the optimal weight combinations varied by dataset—some favored two dominant views, others preferred balanced weighting.  
A fixed-weight model would thus require dataset-specific tuning to remain competitive.  
Because \system{} employs seven monotonic detectors that collectively \emph{cover} representative linear fusion patterns—from single-view to all-view combinations—the voting stage maintains performance comparable to the best observed weights,  and consistently stable across datasets. 
This alignment between theoretical coverage and observed consistency supports the soundness of the overall design.  
% The voting mechanism itself is analyzed in the next subsection (\S\ref{sec:voting_effectiveness}).

\noindent\textbf{Summary.}
Under our uniform experimental setting and datasets:  
(i) monotonic linear fusion provides a reliable basis for integrating multi-view anomaly scores; and  
(ii) the voting-based decision integration mitigates dataset-specific sensitivity, yielding stable results without further tuning.

\vspace{-5pt}
\subsubsection{Ablation: Detector Group Effectiveness and Voting Sensitivity}
\vspace{-5pt}
\label{sec:voting_effectiveness}
Within the multi-dimensional anomaly fusion and voting-based detection framework, \system{} employs three detector groups corresponding to single-, two-, and three-view combinations.  
To assess the contribution of each group and the sensitivity of the voting threshold, we conduct two ablation studies analyzing their respective impacts on detection performance.

\begin{table}
\centering
\begin{small}
\begin{tblr}{
  rowsep = 0pt,
  colsep = 1.2pt,
  cells = {c},
  width = 0.8\linewidth,
  width = \linewidth,
  colspec = {Q[206]Q[98]Q[98]Q[77]Q[119]Q[98]Q[98]Q[58]Q[58]},
  cell{1}{1} = {r=2}{},
  cell{1}{2} = {c=2}{},
  cell{1}{4} = {c=2}{},
  cell{1}{6} = {c=2}{},
  cell{1}{8} = {c=2}{},
  hline{1,10} = {-}{0.1em},
  hline{3,9} = {-}{0.05em},
  hline{2} = {3,5,7,9,11,13,15,17,19}{r},
  hline{2} = {2,4,6,8,10,12,14,16,18}{l},
}

Dataset   & \textit{w/o} Group 1 &    & \textit{w/o} Group 2 &    & \textit{w/o} Group 3 &      & Final &    \\
          & TP                   & FP & TP                   & FP   & TP                   & FP & TP    & FP \\
CADETS-E3 & 24                   & 1  & 24                   & 1    & 24                   & 1  & 24    & 1  \\
THEIA-E3  & 91                   & 2  & 96                   & 214  & 94                   & 5  & 91    & 2  \\
CLRSCP-E3 & 6                    & 7  & 44                   & 1367 & 6                    & 8  & 6     & 7  \\
CADETS-E5 & 4                    & 3  & 7                    & 23   & 8                    & 16 & 7     & 9  \\
THEIA-E5  & 11                   & 2  & 11                   & 32   & 14                   & 41 & 11    & 2  \\
CLRSCP-E5 & 10                   & 16 & 13                   & 80   & 10                   & 18 & 10    & 16 
\end{tblr}
\vspace{-5pt}
\caption{An ablation study of different detector groups. While individual detector groups yield high False Positives or low True Positives, our ensemble method achieves a more effective balance.}
\vspace{-10pt}
\label{tab:detector_remove_group}
\end{small}
\end{table}

\begin{table*}
\centering
\begin{small}
\begin{tblr}{
  rowsep = 0pt,
  colsep = 0.6pt,
  cells = {c},
  width = 1.0\linewidth,
  colspec = {Q[230]Q[35]Q[35]Q[60]Q[42]Q[58]Q[60]Q[58]Q[35]Q[35]Q[60]Q[42]Q[60]Q[60]Q[60]},
  cell{1}{1} = {r=2}{},
  cell{1}{2} = {c=7}{0.348\linewidth},
  cell{1}{9} = {c=7}{0.352\linewidth},
  hline{1,7} = {-}{0.1em},
  hline{3} = {-}{0.05em},
  hline{2} = {2-7}{0.05em},
  hline{2} = {8-15}{0.05em},
  vline{2,9} = {-}{0.05em},
  hline{2} = {8,15}{r},
  hline{2} = {2,9}{l},
}
Ablation Component & Single centric Performance &    &        &     &        &        &        & Final Performance &    &        &     &        &        &        \\
                                                   & TP↑ & FP↓     & TN↑     & FN↓  & F1↑    & ADP↑   & MCC↑    & TP↑ & FP↓     & TN↑     & FN↓  & F1↑    & ADP↑   & MCC↑    \\
Edge-centric$_{w/\ E\text{-}Weight}$                       & 95 & 36 & 701.5K & 34  & 0.73 & 1.00  & 0.73  & 91 & 2  & 701.5K & 38  & 0.82 & 1.00  & 0.83  \\
Edge-centric$_{w/o\ E\text{-}Weight}$                       & 19 & 10 & 701.4K & 110 & 0.24 & 0.73  & 0.31  & 16 & 3  & 701.5K & 113 & 0.22 & 0.64  & 0.32  \\
Node-centric$_{w/\ SA\text{-}Split}$                          & 37 & 19 & 701.5K & 92  & 0.40 & 0.48  & 0.44  & 91 & 2  & 701.5K & 38  & 0.82 & 1.00  & 0.83  \\
Node-centric$_{w/o\ SA\text{-}Split}$                         & 0  & 19 & 701.5K & 129 & 0.00 & 0.00  & 0.00  & 6  & 0  & 701.5K & 123 & 0.09 & 1.00  & 0.22  
\end{tblr}
    \vspace{-5pt}
\caption{Impact of our core detector enhancements on the THEIA-E3 dataset. This ablation study shows that removing either the edge weighting or the structure-attribute (SA) split causes a significant drop in performance.}
    \vspace{-5pt}

\label{tab:enhancement_ablation}
\end{small}
\end{table*}

\begin{table*}
\centering
\begin{small}
\begin{tblr}{
  rowsep = 0pt,
  colsep = 3.2pt,
  cells = {c},
  width = 1.0\linewidth,
  cell{2}{2} = {r=6}{},
  cell{2}{10} = {r=6}{},
  cell{8}{2} = {r=6}{},
  cell{8}{10} = {r=6}{},
  cell{14}{2} = {r=6}{},
  cell{14}{10} = {r=6}{},
  hline{1,14} = {-}{0.1em},
  hline{2,8} = {-}{0.05em},
  vline{2,3,10,11} = {-}{0.05em},
}
Dataset     & System            & TP↑ & FP↓   & TN↑     & FN↓  & F1↑   & ADP↑   & MCC↑   & System            & TP↑ & FP↓   & TN↑     & FN↓  & F1↑   & ADP↑   & MCC↑   \\
CADETS-E3   & \makecell{w/o \\ attribute \\ view} & \textbf{24}  & \textbf{1}     & \textbf{281.5K}  & \textbf{44}   & \textbf{0.52}   & \textbf{1.00}   & \textbf{0.58}   & \makecell{w/o \\ structure \\ view} & \textbf{24}  & \textbf{1}     & \textbf{281.5K}  & \textbf{44}   & \textbf{0.52}   & \textbf{1.00}   & \textbf{0.58}   \\
THEIA-E3    &                   & 90  & \textbf{1}     & \textbf{701.5K}  & 39   & 0.82   & \textbf{1.00}   & 0.83   &                   & 61  & 17    & 701.5K  & 68   & 0.59   & \textbf{1.00}   & 0.61   \\
CLRSCP-E3 &                 & 6   & 288   & 111.1K  & 43   & 0.04   & \textbf{1.00}   & 0.05   &                   & 3   & \textbf{0}     & \textbf{111.4K}  & 46   & 0.12   & \textbf{1.00}   & \textbf{0.25}   \\
CADETS-E5   &                   & 3   & 5     & 3.14M   & 120  & 0.05   & \textbf{1.00}   & 0.10   &                   & 2   & \textbf{3}     & 3.14M   & 121  & 0.03   & 0.90   & 0.08   \\
THEIA-E5    &                   & \textbf{14}  & 35    & 1.86M   & \textbf{55}   & 0.24   & \textbf{1.00}   & 0.24   &                   & 1   & 10    & 1.86M   & 68   & 0.03   & \textbf{1.00}   & 0.04   \\
CLRSCP-E5 &                 & 1   & \textbf{11}    & \textbf{150.9K}  & 52   & 0.03   & \textbf{0.83}   & 0.04   &                   & \textbf{11}  & 23    & 150.9K  & \textbf{42}   & \textbf{0.25}   & 0.82   & 0.26   \\
\hline
CADETS-E3   & \makecell{w/o \\ causal \\ view} & 0   & \textbf{1}     & \textbf{281.5K}  & 68   & 0.00   & 0.85   & 0.00   & \makecell{original \\ ProvFusion} & \textbf{24}  & \textbf{1}     & \textbf{281.5K}  & \textbf{44}   & \textbf{0.52}   & \textbf{1.00}   & \textbf{0.58}   \\
THEIA-E3    &                   & 1   & 15    & 701.5K  & 128  & 0.01   & 0.34   & 0.02   &                   & \textbf{91}  & 2     & 701.5K  & \textbf{38}   & \textbf{0.82}   & \textbf{1.00}   & \textbf{0.83}   \\
CLRSCP-E3 &                 & \textbf{38}  & 983   & 110.4K  & \textbf{11}   & 0.07   & 0.97   & 0.17   &                   & 6   & 7     & 111.3K  & 43   & \textbf{0.19}   & 0.86   & 0.24   \\
CADETS-E5   &                   & 5   & 9     & 3.14M   & 118  & 0.07   & \textbf{1.00}   & 0.12   &                   & \textbf{7}   & 9     & \textbf{3.14M}   & \textbf{116}  & \textbf{0.10}   & 0.92   & \textbf{0.16}   \\
THEIA-E5    &                   & 0   & 12    & 1.86M   & 69   & 0.00   & 0.45   & 0.00   &                   & 11  & \textbf{2}     & \textbf{1.86M}   & 58   & \textbf{0.27}   & \textbf{1.00}   & \textbf{0.37}   \\
CLRSCP-E5 &                 & 4   & 20    & 150.9K  & 49   & 0.10   & 0.76   & 0.11   &                   & 10  & 16    & 150.9K  & 43   & \textbf{0.25}   & 0.77   & \textbf{0.27}   \\
\end{tblr}

    \vspace{-5pt}
\caption{Ablation study quantifying the importance of each analysis view. Removing any single view results in a significant performance drop on multiple datasets, demonstrating effectiveness of our multi-view fusion.}
    \vspace{-10pt}

\label{tab:ablation_views}
\end{small}
\end{table*}

\noindent\textbf{Ablation of Detector Groups.}  
We assess each group’s contribution by disabling it in turn and measuring performance changes across datasets (Table~\ref{tab:detector_remove_group}).  
The results indicate that removing any group degrades performance.  
Excluding \textit{Group 2} or \textit{Group 3} increases false positives, while removing \textit{Group 1} causes a missed attack on CADETS-E5, reducing campaign coverage.  
These observations indicate that each group captures distinct anomaly patterns, and combining all three yields broader coverage with fewer false alarms.  
Detailed per-group and per-detector results are provided in Appendix~\ref{apd:performance_of_each_detector} and Appendix~\ref{apd:performance_of_each_group}.

\noindent\textbf{Threshold sensitivity of the voting rule.}  
An entity is flagged as anomalous if it receives at least $T_v$ votes from the seven detectors.  
We vary $T_v$ from 1 to 7 and record MCC, F1, and false-positive counts (Figure~\ref{fig:threshold_sensitivity}).  
When the threshold is small ($T_v=1$–2), the detection criterion becomes overly permissive, flagging many benign nodes as anomalies and resulting in a sharp increase in false positives.  
Conversely, when the threshold is large ($T_v=6$–7), the detection criterion becomes too strict: although false positives drop, true positives also decline noticeably, leading to lower F1 and MCC scores.  
Between these extremes, moderate thresholds ($T_v=3$–5) maintain high F1 and MCC values while keeping false positives at a manageable level, indicating a balanced trade-off between recall and precision.  
The default $T_v=4$ (a simple majority of seven) lies within this stable range and provides consistently strong results across datasets.

\noindent\textbf{Summary.}  
Each detector group contributes distinct strengths, and the voting mechanism remains robust within a wide threshold range ($T_v=3$–5), showing consistent behavior across datasets.

\vspace{-5pt}
\subsubsection{Ablation: Enhancements to Node- and Edge-Centric Detectors}
\vspace{-5pt}
\label{sec:detector_enhancement}
\system{} \emph{incorporates} two view-specific enhancements as part of its architecture. 
(i) For the \textbf{node-centric} view, we decouple structural and attribute-based scoring to mitigate false positives caused by noisy or infrequent attributes (e.g., rare file paths). 
(ii) For the \textbf{edge-centric} view, we apply a class-weighted training loss to address the imbalance between frequent and rare edge types in provenance graphs. 
To assess their contribution \emph{within \system{}}, we perform ablations by removing each enhancement in turn and measuring the impact on detection performance.

\noindent\textbf{Experimental setup.}  
We compare the full system (Edge$_{w/\ E\text{-}Weight}$, Node$_{w/\ SA\text{-}Split}$) with two degraded variants:  
Node$_{w/o\ SA\text{-}Split}$, which concatenates one-hot node types with attribute embeddings as input features, causing the encoder to aggregate structural and attribute information into a single representation; and 
Edge$_{w/o\ E\text{-}Weight}$, which treats all edge types equally.  
Results on THEIA-E3, a representative dataset, are shown in Table~\ref{tab:enhancement_ablation}.

\noindent\textbf{Findings.}  
Introducing the structural–attribute split increases true positives, improving the system’s ability to identify real attack nodes while filtering attribute-level noise.  
Disabling edge weighting lowers ADP, indicating reduced ability to emphasize rare but critical edges relative to common benign ones.  
Both enhancements therefore strengthen the detectors’ discriminative power and stability.  
% We next examine how these enhanced view-specific detectors cooperate within the multi-view fusion framework (\S\ref{sec:multiview_analysis}).

\vspace{-5pt}
\subsubsection{Contribution of the Multi-View Architecture}
\vspace{-5pt}
\label{sec:multiview_analysis}
To assess the contribution of each view, we disable one at a time and re-evaluate performance (Table~\ref{tab:ablation_views}).  
All three views contribute meaningfully but in distinct ways.  
Removing the \textit{Edge-centric Causal View} mainly reduces recall, whereas removing either the \textit{Node-centric Attribute View} or the \textit{Structure View} increases false positives and lowers precision.  
These patterns indicate that each view captures distinct aspects of anomalous behavior, and together they contribute to balanced and stable detection across datasets.

\vspace{-5pt}
\begin{tcolorbox}[colback=gray!5!white, colframe=gray!70!black, boxrule=0.7pt, arc=3pt, width=\linewidth]
\textbf{RQ2 Answer.} 
Our in-depth analysis shows that \system{}’s detection capability results from the coordinated contributions of multiple design components rather than any single factor.  
(1) Enhancements to the node- and edge-centric detectors reinforce the underlying anomaly signals;  
(2) the multi-dimensional anomaly fusion with voting-based detection helps achieve stable performance and robustness to moderate threshold variations;
 and  
(3) the multi-view architecture allows different types of anomalies to be captured across diverse scenarios.  Overall, \system{}’s effectiveness arises from the synergy of these components, which together provide reliable and consistent detection across datasets.

\end{tcolorbox}
\vspace{-5pt}

\vspace{-9pt}
\subsection{Computational Overhead (RQ3)}
\vspace{-9pt}
\label{sec:overhead}

In this section, we evaluate the runtime and scalability of \system{} to answer our third research question:  
\textit{Is \system{} computationally efficient and scalable?}  
We measure the training time, inference time, and scalability with respect to graph size, and compare \system{} with Velox~\cite{velox} and Orthrus~\cite{orthrus}, two SOTA systems that report higher efficiency than all other compared approaches. 
All experiments are conducted under the same hardware and software configuration for fair comparison.

\begin{figure}[h]
  \centering
  \begin{minipage}{0.49\linewidth}
    \centering
    \includegraphics[width=\linewidth]{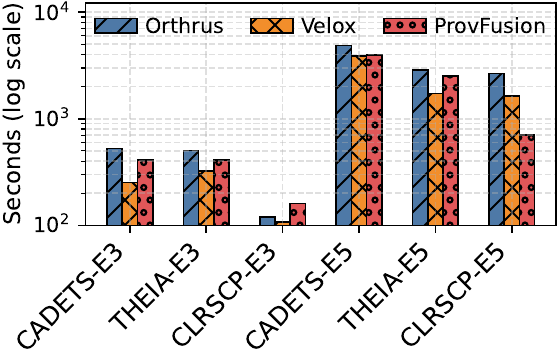}
    \vspace{-15pt}
    \caption*{(a) Training Time.}
    \vspace{-5pt}
    \label{fig:train_time_a}
  \end{minipage}
  \hfill
  \begin{minipage}{0.49\linewidth}
    \centering
    \includegraphics[width=\linewidth]{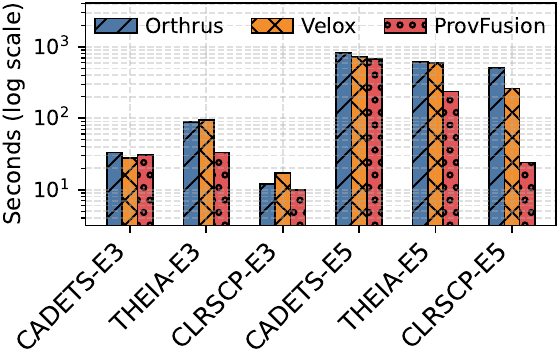}
    \vspace{-15pt}
    \caption*{(b) Inference Time.}
    \vspace{-5pt}
    \label{fig:test_time_a}
  \end{minipage}
  \caption{Runtime Overhead.}
  \vspace{-5pt}
  \label{fig:rumtime}
\end{figure}

\vspace{-9pt}
\subsubsection{Training and Inference Efficiency}

Figure~\ref{fig:rumtime} shows the training and inference times across datasets, with a detailed runtime breakdown 
in Appendix~\ref{app:runtime_breakdown}.
Overall, \system{} achieves broadly comparable training efficiency to Velox and Orthrus, while exhibiting higher \emph{inference} efficiency, as we detail below.

First, during both training and inference, \system{} represents multiple edges between two nodes as a single multi-hot edge, rather than processing each edge individually. This design avoids redundant computation and reduces the total number of edge-level operations, thereby mitigating part of the additional cost introduced by the multi-view framework. As a result, the overall training time is competitive. For instance, on the largest dataset, CADETS-E5, our training time is comparable to Velox (with only $\sim$2.6\% overhead) and faster than Orthrus.

Second, a critical component of \system{} is the KNN-based scoring for the attribute and structure views. While this step, which compares test nodes against all training nodes, is often a computational bottleneck, we address this by integrating the Faiss library~\cite{douze2024faiss,johnson2019billion} to accelerate the nearest-neighbor search. This optimization is highly effective, leading to a consistently strong inference performance. As a result, \system{} is faster than both baselines on many datasets. For instance, on the largest dataset (CADETS-E5), \system{} is $\sim$1.10$\times$ faster than Velox and $\sim$1.25$\times$ faster than Orthrus. On THEIA-E5, it is $\sim$2.5$\times$ faster than both baselines, demonstrating the practical efficiency of our optimized framework.

\begin{figure}[t]
  \centering
  \begin{minipage}{0.49\linewidth}
    \centering
    \includegraphics[width=\linewidth]{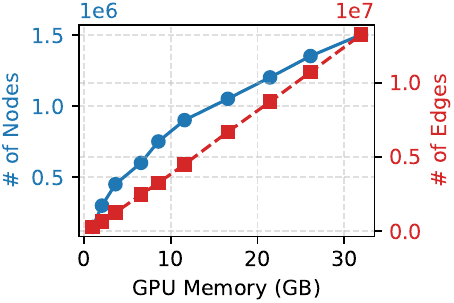}
    \vspace{-15pt}
    \caption*{(a) Peak GPU Memory.}
    \vspace{-5pt}
    \label{fig:train_time_b}
  \end{minipage}
  \hfill
  \begin{minipage}{0.49\linewidth}
    \centering
    \includegraphics[width=\linewidth]{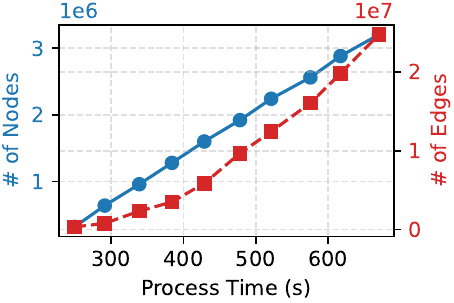}
    \vspace{-15pt}
    \caption*{(b) Inference Throughput.}
    \vspace{-5pt}
    \label{fig:test_time_b}
  \end{minipage}
  \caption{Scalability Analysis.}
  \vspace{-15pt}
  \label{fig:scalability}
\end{figure}

\vspace{-9pt}
\subsubsection{Scalability Analysis}

To evaluate scalability, we construct graphs of increasing size from CADETS-E5 by randomly sampling nodes (300K to 3M) and their associated edges. We then measure GPU memory usage and inference processing time against the graph size.
As shown in Figure~\ref{fig:scalability}, GPU memory consumption scales approximately linearly with the graph size, as expected.
More importantly, the inference time exhibits a favorable \textbf{linear} growth along with graph size. Our analysis shows that while the number of nodes increases by 10$\times$ (from 3K to 3M) and the number of edges increases by $\sim$75$\times$ (from 330K to 24.7M), the total processing time increases by only $\sim$2.7$\times$ (from 249s to 670s). This result demonstrates that our Faiss-accelerated~\cite{douze2024faiss} KNN scoring step successfully avoids becoming a quadratic bottleneck.

\vspace{-5pt}
\begin{tcolorbox}[colback=gray!5!white, colframe=gray!70!black, boxrule=0.7pt, arc=3pt, width=\linewidth]
\vspace{-5pt}
\textbf{RQ3 Answer.}
\system{} achieves competitive training efficiency and relatively faster inference performance compared to SOTA baselines, particularly on large-scale datasets. This efficiency is driven by our multi-hot edge representation and Faiss-accelerated KNN scoring. Scalability analysis confirms that the memory usage and the inference processing time grow linearly with graph size. This demonstrates that our framework is scalable and efficient for large-scale graphs.
\vspace{-5pt}
\end{tcolorbox}
\vspace{-5pt}

\begin{figure*}[t]
    \centering
    \includegraphics[width=0.95\textwidth]{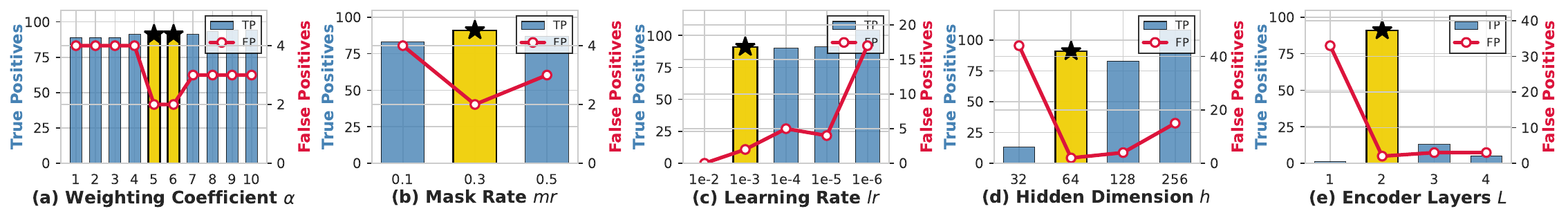}
    \vspace{-8pt}
    \caption{Hyperparameter sensitivity on THEIA-E3 dataset.}
    \vspace{-10pt}
    \label{fig:hyperparam}
\end{figure*}

\vspace{-5pt}
\subsection{Hyperparameter Study (RQ4)}
\vspace{-5pt}
\label{sec:hyperparameter}

In this section, we answer RQ4 using the THEIA-E3 dataset, which contains a relatively large number of labeled nodes (128) and three distinct attack campaigns, making it suitable for controlled sensitivity analysis.  
Figure~\ref{fig:hyperparam} summarizes the results, reporting true positives (TPs), false positives (FPs) under different parameter settings.

\noindent\textbf{Weighting coefficient $\alpha$.}
As described in Section~\S\ref{sec:fusion_detection}, $\alpha$ controls the adaptive weighting.
Performance remains highly consistent for $\alpha$ between 4 and 7 (TPs stay at 91, FPs at 2--4), showing limited sensitivity to this parameter.

\noindent\textbf{Mask rate $mr$.}
A moderate rate of $mr=0.3$ yields the better balance (91 TPs, 2 FPs), while lower ($mr=0.1$) and higher ($mr=0.5$) rates degrade performance (83 TPs / 4 FPs and 87 TPs / 3 FPs, respectively).

\noindent\textbf{Learning rate $lr$.}
A proper learning rate ensures smooth training dynamics.
We find that $lr=1\times10^{-3}$ provides the best performance (91 TPs, 2 FPs), as higher rates (e.g., $1e-2$) failed to converge (0 TPs), and lower rates produced more false positives (e.g., 5 FPs at $1e-4$ or 17 FPs at $1e-6$).

\noindent\textbf{Encoder layers $L$.}
The number of GNN encoder layers controls neighborhood aggregation.
A two-layer ($L=2$) encoder achieves the best balance (91 TPs, 2 FPs).
A single layer ($L=1$) provides insufficient representational capacity (1 TP, 33 FPs), while too many layers (e.g., $L=3$ or $L=4$) cause over-smoothing, degrading performance (13 TPs and 5 TPs, respectively).

\noindent\textbf{Hidden dimension $h$.}
The hidden dimension determines the representational granularity.
A dimension of $h=64$ provides an effective balance (91 TPs, 2 FPs). A smaller dimension ($h=32$) underfits and performs poorly (13 TPs, 44 FPs), while a larger one ($h=128$) slightly degrades performance (83 TPs, 4 FPs).

\vspace{-5pt}
\begin{tcolorbox}[colback=gray!5!white, colframe=gray!70!black, boxrule=0.7pt, arc=3pt, width=\linewidth]
\vspace{-5pt}
\textbf{RQ4 Answer.}  
\system{} shows stable performance across its core tuning parameters, with limited sensitivity to the fusion coefficient $\alpha$ and mask rate $mr$.  
Consistent with general GNN behavior, architectural parameters such as encoder layers ($L$) and hidden dimension ($h$) require careful tuning to balance representational capacity and avoid underfitting or over-smoothing.
\vspace{-5pt}
\end{tcolorbox}
\vspace{-5pt}

\vspace{-5pt}
\section{Adversarial Robustness}
\vspace{-5pt}
\label{subsec:adversarial_attacks}

A representative example is the mimicry attack proposed by Goyal et al.~\cite{sometimes}, which targets \textit{subgraph-based} detectors (e.g., Unicorn~\cite{unicorn}) by crafting malicious subgraphs that imitate benign topological motifs. These attacks rely on the assumption of \textit{global structural similarity}, embedding malicious interactions within subgraphs that visually and statistically resemble benign patterns. However, this assumption does not directly apply to node- or edge-centric paradigms, which assess anomalies based on localized behavioral contexts—focusing on entity-level attributes and interaction plausibility rather than global subgraph resemblance.
This distinction has been empirically supported by the benchmark analysis of Velox~\cite{velox}, which evaluated multiple recent PIDSs~\cite{nodlink,threatrace,kairos,velox,orthrus,magic,flash} under subgraph-level mimicry attacks and found that node- and edge-centric systems were largely unaffected, as localized behavioral features remained stable under such global perturbations. Consequently, mimicry attacks designed for subgraph-based detectors are not directly suitable for evaluating systems following node- or edge-centric paradigms such as ours.

To empirically assess the reasoning, we reproduced the mimicry attack experiment of Goyal et al.~\cite{sometimes} using their publicly released implementation~\cite{velox, sometimes}. Detailed results are in Appendix~\S\ref{sec:appendix_robustness}. Consistent with prior observations, the attack failed to induce false negatives in \system{}, and all original malicious entities were correctly identified. However, we observed a subtle secondary effect: the structural perturbations introduced by the attack slightly increased false positives. This likely occurred because the edges inserted by the attacker—intended to make the subgraph appear benign—introduced local behavioral inconsistencies that our fine-grained detectors recognized as anomalous.

This finding provides empirical confirmation that subgraph-level mimicry attacks are largely ineffective against node- and edge-centric paradigms, whose decisions rely on localized rather than global behavioral evidence. It also points to a next-step challenge—designing adaptive evasion strategies that can deceive fine-grained detectors without introducing new detectable inconsistencies, thereby enabling a more rigorous evaluation of PIDSs.
\vspace{-5pt}
\section{Related Work}
\vspace{-4pt}
\label{sec:related_work}

We position \system{} with respect to two established paradigms in provenance-based intrusion detection: \emph{node-centric} and \emph{edge-centric} analysis. Each paradigm provides a distinct analytical focus. Reviewing them helps clarify the scope of prior progress and the rationale for combining multiple analytical perspectives.

\noindent\textbf{Node-centric approaches.}
Node-centric methods analyze the properties of individual entities and their immediate neighborhoods. They generally operate under the notion that malicious behavior alters the attributes or structural context of affected nodes.  
Representative works such as ThreaTrace~\cite{threatrace} and Flash~\cite{flash} train GNNs to predict node types (e.g., process, file) based on contextual information, with classification inconsistency used as an indicator of potential abnormality.  
Other methods adopt self-supervised objectives to model regularities of benign behavior without relying on predefined labels.  
SIGL~\cite{sigl} employs a Graph LSTM-based~\cite{graphlstm} autoencoder to reconstruct process features, while MAGIC~\cite{magic} uses a GAT-based masked reconstruction framework and measures deviations via search-based metrics such as K-D tree distance~\cite{knn}.  
R-CAID~\cite{rcad} further combines GNNs with root-cause reasoning to associate detected deviations with their potential origins.  
Overall, these approaches provide detailed per-entity assessment but remain centered on local reasoning. When coordinated attack behaviors span multiple entities that each appear individually plausible, purely node-centric detectors may provide limited visibility into the broader execution context.

\noindent\textbf{Edge-centric approaches.}
Edge-centric methods~\cite{shadewatcher} shift the analytical focus from entities to their interactions, modeling whether interactions between entities conform to typical causal patterns.  
Kairos~\cite{kairos} leverages Temporal Graph Networks (TGNs)~\cite{tgn} to represent long-range temporal dependencies in system event streams.  
To improve scalability, Orthrus~\cite{orthrus} and Velox~\cite{velox} introduce more lightweight formulations emphasizing efficiency and interpretability.  
Such systems are effective at identifying irregular causal flows but may miss anomalies residing within entity states.  
For instance, a compromised process may continue to perform syntactically valid operations, generating causal links that appear ordinary when viewed in isolation.  
In such cases, evaluating edge plausibility alone may not sufficiently expose the underlying compromise.

\noindent\textbf{Synthesis.}
Existing work collectively highlights a persistent granularity gap.  
Node-centric detectors focus primarily on entity states and tend to place limited emphasis on interaction semantics, whereas edge-centric detectors concentrate on causal relationships while offering only partial visibility into the internal conditions of participating entities.  
Both perspectives contribute valuable insights yet capture different aspects of system behavior.  
As modern attack campaigns often involve subtle changes across entities and their interactions, reliance on a single detection view can reduce coverage in complex scenarios.  
\system{} is designed to address this practical limitation by jointly analyzing entity-level and interaction-level anomalies within a unified multi-view formulation, providing a more context-aware basis for provenance-based intrusion detection.

\vspace{-6pt}
 \rev{\section{Limitations and Future Work}
 \label{sec:limitations}}
\vspace{-8pt}

% \smallskip
\rev{\noindent\textbf{Static graph and temporal dynamics.}
\system{} encodes the events between an entity pair as a single multi-hot edge, which discards their ordering and timing. Like other static-graph PIDSs, it may therefore underestimate attacks that are anomalous only in sequence (e.g., slow exfiltration interleaved with legitimate accesses). Integrating temporal modeling such as temporal graph networks~\cite{tgn} into the multi-view pipeline is a promising direction.}

% \smallskip
\rev{\noindent\textbf{Single-host scope.} \system{}, like existing PIDSs~\cite{magic,nodlink,flash,velox,orthrus,kairos}, analyzes provenance from a single host and does not directly capture lateral movement across hosts. Extending multi-view fusion to cross-host settings requires addressing heterogeneous logging formats, cross-boundary entity resolution, and substantially larger graph scales; we view federated or hierarchical fusion across host-level detectors as a natural next step.}

% \smallskip
\rev{\noindent\textbf{Noisy and incomplete logging.} We assume faithful provenance capture (Section~\S\ref{sec:threat_model}); in practice, log pruning failures can break causal chains and degrade the structural and causal views. Developing uncertainty-aware scoring or imputation under degraded logging is a valuable future direction.}

% \vspace{3pt}
\section{Conclusion}
\vspace{-8pt}
In this work, we examined the different focuses of node-centric and edge-centric approaches in provenance-based intrusion detection. To bridge these gaps, we introduced \system{}, a multi-view detection framework that integrates anomalies from three detection views: entity attributes, structure pattern, and causal interactions. By systematically fusing these views and employing a voting-based detection mechanism that relies solely on validation data, \system{} provides a broader and more consistent assessment of system behavior. Experiments on six benchmark datasets show improved detection quality over representative state-of-the-art systems, with higher detection rates and lower false positives. Overall, this work offers an effective design for provenance-based defense systems.
\vspace{-5pt}
\let\OLDthebibliography\thebibliography
\renewcommand\thebibliography[1]{
  \OLDthebibliography{#1}
  \setlength{\parskip}{0pt}
  \setlength{\itemsep}{0pt}
}
\bibliographystyle{IEEEtran}
\bibliography{ref}

\setcounter{section}{0}
\renewcommand{\thesection}{\Alph{section}}

\newcommand{\appsection}[2]{%
  \refstepcounter{section}% 步进 section 计数器（会创建可引用锚点）
  \section*{\thesection.\ #1}% 视觉上的标题
  \addcontentsline{toc}{section}{Appendix \thesection.\ #1}% 进目录
  \label{#2}% 现在 #2 可被 \ref 正确引用为 A/B/…
}

\vspace{-8pt}
\appsection{Entity and Event Type Specification}{apd:entity_event_types}
\vspace{-10pt}

Table~\ref{tab:entity_event_types} enumerates the entity and event types used in constructing the provenance graph $G=(V,E)$.

\noindent

\begin{table}[h]
\centering
{
% \color{diffblue}
% \arrayrulecolor{diffblue}
\begin{tabular}{cc}
\toprule
\textbf{Category} & \textbf{Types Considered in \system} \\
\midrule
\textbf{Entity Types} & \texttt{process}, \texttt{file}, \texttt{netflow} \\
\textbf{Event Types} & 
\begin{tabular}[t]{@{}l@{}}
\texttt{CONNECT}, 
\texttt{EXECUTE},  
\texttt{OPEN},  
\texttt{READ}, 
\texttt{RECVFROM},  \\
\texttt{RECVMSG},  
\texttt{SENDMSG},  
\texttt{SENDTO},  
\texttt{WRITE}, 
\texttt{CLONE}
\end{tabular} \\
\bottomrule
\end{tabular}
}
\vspace{-5pt}
% \captionsetup{labelfont={color=diffblue}, textfont={color=diffblue}}
\caption{Summary of entity and event types.}
\label{tab:entity_event_types}
\vspace{-10pt}
\end{table}

\vspace{-10pt}
\appsection{Dataset Statistics}{apd:statistics}
\vspace{-10pt}
Table~\ref{tab:dataset_statistics} presents detailed statistics for each of the six processed DARPA TC datasets.

\begin{table}[h]
\centering
\begin{small}
\begin{tblr}{
  rowsep = -0.5pt,
  colsep = 1.2pt,
  cells = {c},
  width = \linewidth,
  colspec = {Q[313]Q[304]Q[302]},
  hline{1,11} = {-}{0.1em},
  hline{2} = {-}{0.05em},
}
Dataset~      & Number of Nodes & Number of Edges \\
CADETS-E3     & 1,103,333         & 4,476,834         \\
THEIA-E3      & 1,308,080          & 5,969,280         \\
ClEARSCOPE-E3 & 260,700          & 838,876          \\
CADETS-E5     & 8,045,589         & 69,319,401        \\
THEIA-E5      & 4,390,381        & 46,035,572        \\
ClEARSCOPE-E5 & 366,025          & 4,523,379        \\
{OpTC-H201} & {3,839,000} & {8,289,319}\\
{OpTC-H501} & {3,884,805} & {8,454,821}\\
{OpTC-H051} & {3,089,126} & {8,182,521}\\
\end{tblr}
\end{small}
\vspace{-5pt}
\caption{Statistics of the evaluation datasets.}
\label{tab:dataset_statistics}
\vspace{-10pt}
\end{table}

\vspace{-8pt}
\appsection{Dataset Splitting Methodology}{apd:dataset_splitting}
\vspace{-10pt}
We adopted the identical data splitting methodology used in recent works~\cite{orthrus, velox}. Table~\ref{tab:apd_data_split} provides the specific date ranges used to partition each dataset. The training sets consist exclusively of benign data, while the test sets contain a mix of benign and malicious activities.

\begin{table}[h]
\begin{small}
\centering
\begin{tblr}{
  rowsep = 0pt,
  colsep = 1.2pt,
  width = \linewidth,
  colspec = {Q[210]Q[300]Q[182]Q[305]},
  row{1} = {c},
  hline{1,8} = {-}{0.08em},
  hline{2} = {-}{0.05em},
  % === 新增：每行之间的横线（与你原来粗细一致）===
  hline{3,4,5,6,7,8,9,10,11} = {-}{0.05em},
  % === 新增：所有垂直线（与你原来水平线粗细协调）===
  vline{1-Z} = {1}{-}{0.08em},
  vline{2,3,4} = {-}{0.05em},
  vline{Z} = {-}{0.08em},
}
\textbf{Datasets}      & \textbf{Train}                         & \textbf{Valid}      & \textbf{Test}                \\
CADETS-E3     & 2018-04-03 $\sim$ 10  & 2018-04-02 & {\scriptsize 2018-04-\textbf{06}/11/\textbf{12}/\textbf{13}} \\
THEIA-E3      & 2018-04-02 $\sim$ 08  & 2018-04-09 & 2018-04-\textbf{10}/\textbf{12}/\textbf{13}    \\
CLRSCP-E3 & 2018-04-03 $\sim$ 10 & 2018-04-02 & 2018-04-\textbf{11}/12       \\
CADETS-E5     & 2019-05-08/09/11              & 2019-05-12 & 2019-05-\textbf{16}/\textbf{17}       \\
THEIA-E5      & 2019-05-08/09/10              & 2019-05-11 & 2019-05-14/\textbf{15}       \\
CLRSCP-E5 & 2019-05-08/09                 & 2019-05-11 & 2019-05-14/\textbf{15}/\textbf{17} \\
{OpTC-H201} & {2019-09-19 $\sim$ 21} & {2019-09-22} & {2019-09-\textbf{23 $\sim$ 25}} \\
{OpTC-H501} & {2019-09-19 $\sim$ 21} & {2019-09-22} & {2019-09-\textbf{23 $\sim$ 25}} \\
{OpTC-H051} & {2019-09-19 $\sim$ 21} & {2019-09-22} & {2019-09-\textbf{23 $\sim$ 25}} \\
\end{tblr}
\vspace{-5pt}
\caption{Date-based partitioning of each dataset in (yyyy-mm-dd). Dates containing malicious attack scenarios in the test set are marked in \textbf{bold}.}
\label{tab:apd_data_split}
\vspace{-8pt}
\end{small}
\end{table}

\vspace{-8pt}
\appsection{Timestamp and Name of Each attack}{apd:detailed_name_and_timea_of_attack}
\vspace{-10pt}
This section illustrates the detailed name and the timestamp of their occurrence. Table~\ref{tab:attack_description} shows the details.

\begin{table}[h]
\begin{small}
\centering
\begin{tblr}{
  rowsep = -0.5pt,
  colsep = 0.8pt,
  cells = {c},
  width = \linewidth,
  % --- 修改在这里 ---
  % 我为所有4列都指定了宽度，并从第3列(652)匀了50个单位给第4列(150)
  colspec = {Q[150]Q[90]Q[652]Q[150]},
  % ----------------
  cell{2}{1} = {r=3}{},
  cell{5}{1} = {r=3}{},
  cell{9}{1} = {r=2}{},
  cell{12}{1} = {r=4}{},
  hline{1,19} = {-}{0.1em},
  hline{2,5,8-9,11-12,16} = {-}{0.05em},
}
% --- 修改在这里 ---
Dataset & Name & Description & Date(yy-mm-dd) \\
% ----------------
CADETS E3 & A1 & Nginx Backdoor w/ Drakon In-Memory & 18-04-06 \\
          & A2 & Nginx Backdoor w/ Drakon In-Memory & 18-04-12 \\
          & A3 & Nginx Backdoor w/ Drakon In-Memory & 18-04-13 \\
THEIA E3 & A1 & Firefox Backdoor w/ Drakon In-Memory & 18-04-10 \\
         & A2 & Browser Extension w/ Drakon Dropper & 18-04-12 \\
         & A3 & Phishing E-mail w/ EXE Attachment & 18-04-13 \\
CLRSCP E3 & A1 & Firefox Backdoo w/ Drakon In-Memory & 18-04-11 \\
CADETS E5 & A1 & Nginx Drakon APT & 19-05-16 \\
          & A2 & Nginx Drakon APT & 19-05-17 \\
THEIA E5 & A1 & Firefox Drakon APT BinFmt-Elevate Inject & 19-05-15 \\
CLRSCP E5 & A1 & Appstarter APK Micro APT Elevate & 18-05-15 \\
          & A2 & Firefox Drakon APT & 19-05-17 \\
          & A3 & Lockwatch APK Java APT & 19-05-17 \\
          & A4 & Tester Micro APT BinFmt-Elevate & 19-05-17 \\
{OpTC} & {A1} & {Attack on OpTC-H201} & {19-09-23} \\
{OpTC} & {A2} & {Attack on OpTC-H501} & {19-09-24} \\
{OpTC} & {A3} & {Attack on OpTC-H051} & {19-09-25} \\
\end{tblr}
\vspace{-5pt}
\caption{Mapping of shorthand attack IDs (A1, A2, …) to their full descriptions.}
\vspace{-10pt}
\label{tab:attack_description}
\end{small}
\end{table}

\vspace{-8pt}
\appsection{
The performance of each detector group.}{apd:performance_of_each_group}
\vspace{-10pt}
We further evaluate the performance of each of the three detector groups in isolation. Table~\ref{tab:group_analysis} presents their standalone performance on all datasets. The results starkly illustrate the necessity of an ensemble approach. No single detector group achieves a satisfactory balance between detection rate and false positives. For instance, while \textit{Group 2} and \textit{Group 3} (i.e., Pairwise Corroboration detector group and Holistic Fusion detector group) identify most attacks, they suffer from an increased number of False alarms. Conversely, \textit{Group 1} (i.e., a Specialist detector group) generates fewer false alarms but fails to detect several stealthy attacks entirely. \textit{This demonstrates that relying on any single detector group, or a simple fusion node-centric and edge-centric method, is insufficient, motivating our ensemble design. }

\begin{table}
\vspace{-10pt}
\centering
\begin{small}
\begin{tblr}{
  rowsep = -0.5pt,
  colsep = 1.2pt,
  cells = {c},
  width = \linewidth,
  colspec = {Q[42]Q[42]Q[42]Q[85]Q[123]Q[142]Q[85]Q[104]Q[104]Q[142]},
  cell{1}{1} = {c=3}{0.126\linewidth},
  cell{1}{4} = {c=2}{0.208\linewidth},
  cell{1}{6} = {c=2}{0.226\linewidth},
  cell{1}{8} = {c=2}{0.208\linewidth},
  cell{1}{10} = {r=2}{},
  hline{1,6} = {-}{0.1em},
  hline{3} = {-}{0.05em},
  hline{2} = {2-3}{0.05em},
  hline{2} = {3,5,7,9,11,13,15,17,19}{r},
  hline{2} = {1,4,6,8,10,12,14,16,18}{l},
}
Groups &   &   & CADETS &        & THEIA   &      & CLRSCP &       \\
1      & 2 & 3 & E3     & E5     & E3      & E5   & E3     & E5    \\
\solidcircle       &   &   & 0/0    & 0/0    & 3/7     & 5/8  & 5/0    & 6/1   \\
       &\solidcircle&   & 24/1   & 94/5   & 6/8     & 8/16 & 14/41  & 10/18 \\
       &   & \solidcircle  & 24/1   & 104/36 & 46/5894 & 4/17 & 11/32  & 13/85 \\
\end{tblr}
\vspace{-5pt}
\caption{Analysis of detector groups with cell of (TP/FP).}
\vspace{-10pt}
\label{tab:group_analysis}
\end{small}
\end{table}

\vspace{-8pt}
\appsection{Ablation Study on Normalization Strategy}{apd:norm_ablation}
\vspace{-10pt}
% \vspace{-5pt}
 The aggregate results are presented in Table~\ref{tab:norm_ablation_summary}. \rev{Across both the TC and OpTC families,} only Percentile Normalization provides the necessary balance of sensitivity and precision required by our framework.

\rev{%
    % \vspace{-8pt}
  \appsection{Benign-Shift Stability Study}{apd:benign_shift}
  \vspace{-8pt}
  To evaluate robustness to a benign distribution shift, we rotate which day
  serves as the validation split on THEIA-E3 (four days, each taking one
  turn as validation). Results are in Table~\ref{tab:benign_shift}.
  \system{} detects all three attack campaigns in every split, confirming that percentile-based fusion is
  not sensitive to the choice of a benign holdout.}

\begin{table}[h]
\centering
\vspace{-5pt}
% \captionsetup{labelfont={color=diffblue}, textfont={color=diffblue}}
\small
\setlength{\tabcolsep}{0pt}
\setlength{\tabcolsep}{4pt}
\setlength{\aboverulesep}{0pt}
\setlength{\belowrulesep}{0pt}
\renewcommand{\arraystretch}{0.92}
% \arrayrulecolor{diffblue}
% \color{diffblue}
\begin{tabular}{llrrrrrr}
\toprule
Training&Validation & TP & FP & F1 & ADP & MCC & Det. \\
\midrule
Days 3,4,5&Day 9 &  91 &  2 & 0.82 & 1.00 & 0.83 & 3/3 \\
Days 4,5,9&Day 3 & 110 & 12 & 0.88 & 1.00 & 0.88 & 3/3 \\
Days 3,5,9&Day 4 &  86 & 14 & 0.73 & 1.00 & 0.74 & 3/3 \\
Days 3,4,9&Day 5 &  95 &  3 & 0.84 & 1.00 & 0.84 & 3/3 \\
\bottomrule
\end{tabular}
\color{black}
\arrayrulecolor{black}
\vspace{-5pt}
\caption{Benign-shift study on THEIA-E3.}
\label{tab:benign_shift}
\vspace{-5pt}
\end{table}

\begin{table}[t]
\centering
\small
\begin{tblr}{
  rowsep = -0.5pt,
  colsep = 1.2pt,
  cells = {c},
  width = \linewidth,
  colspec = {Q[120]Q[260]Q[160]Q[85]Q[117]Q[181]},
  column{even} = {c},
  column{4} = {c},
  column{6} = {c},
  hline{1,10} = {-}{0.1em},
  hline{2,6} = {-}{0.05em},
}
\textbf{Dataset} & \textbf{Normalization}     & \textbf{Detection} & \textbf{TPs} & \textbf{ADP}   & \textbf{Total FPs} \\
TC   & \textbf{Percentile} & \textbf{13/14}     & \textbf{149} & \textbf{0.926} & \textbf{37}        \\
TC   & Min-Max Scaling            & 8/14               & 110          & 0.848          & 2378               \\
TC   & Z-Score Scaling            & 5/14               & 18           & 0.791          & 27                 \\
TC   & Robust Scaling             & 1/14               & 6            & 0.650          & 144                \\
{OpTC} & {\textbf{Percentile}} & {\textbf{3/3}} & {\textbf{21}} & {\textbf{0.980}} & {\textbf{39}} \\
{OpTC} & {Min-Max Scaling}            & {2/3}          & {2}           & {0.786}          & {25}          \\
{OpTC} & {Z-Score Scaling}            & {0/3}          & {0}           & {0.223}          & {19}          \\
{OpTC} & {Robust Scaling}             & {2/3}          & {5}           & {0.897}          & {144}         \\
\end{tblr}
\vspace{-5pt}
% \captionsetup{labelfont={color=blue},textfont={color=blue}}
\caption{Aggregate Performance of Normalization Strategies on TC and OpTC Datasets.}
\label{tab:norm_ablation_summary}
\vspace{-3pt}
\end{table}

\begin{table}
\centering
\begin{small}
\begin{tblr}{
  rowsep = -0.5pt,
  colsep = 1.2pt,
  cells = {c},
  width = \linewidth,
  colspec = {Q[179]Q[92]Q[133]Q[133]Q[112]Q[152]Q[112]},
  cell{1}{1} = {r=2}{},
  cell{1}{2} = {c=2}{0.225\linewidth},
  cell{1}{4} = {c=2}{0.245\linewidth},
  cell{1}{6} = {c=2}{0.264\linewidth},
  vline{2,4,6} = {1-10}{0.05em},
  hline{1,11} = {-}{0.1em},
  hline{3} = {-}{0.05em},
}
\textbf{Detectors} & \textbf{CADETS} &        & \textbf{THEIA}  &       & \textbf{CLRSCP}  &       \\
          & \textbf{E3}     & \textbf{E5}     & \textbf{E3}     & \textbf{E5}    & \textbf{E3}      & \textbf{E5}    \\
D1        & 0/67   & 14/249 & 0/25   & 7/21  & 38/2240 & 0/23  \\
D2        & 0/2    & 6/314  & 1/3882 & 0/30  & 3/522   & 9/23  \\
D3        & 24/1   & 0/15   & 95/36  & 12/47 & 6/8     & 8/14  \\
D4        & 24/1   & 8/16   & 91/2   & 14/41 & 6/7     & 10/18 \\
D5        & 25/2   & 8/16   & 94/5   & 14/41 & 6/8     & 10/18 \\
D6        & 24/1   & 4/17   & 91/360 & 1/31  & 43/2445 & 6/79  \\
D7        & 24/1   & 4/17   & 94/7   & 11/32 & 9/3929  & 13/85 \\
All       & 24/1   & 7/9    & 91/2   & 11/2  & 6/7     & 10/16 
\end{tblr}
\vspace{-5pt}
\caption{A comparison of individual detectors (D1–D7) and the final fused output with each cell TP/FP.}
\vspace{-10pt}
\label{tab:detector_eac}
\end{small}
\end{table}

\vspace{-8pt}
\appsection{Case Study: Reconstructing an Attack from Sparse Signals}{apd:case_study}
\vspace{-10pt}
\begin{figure}[h]
\centering
\vspace{-5pt}
\includegraphics[width=0.9\columnwidth]{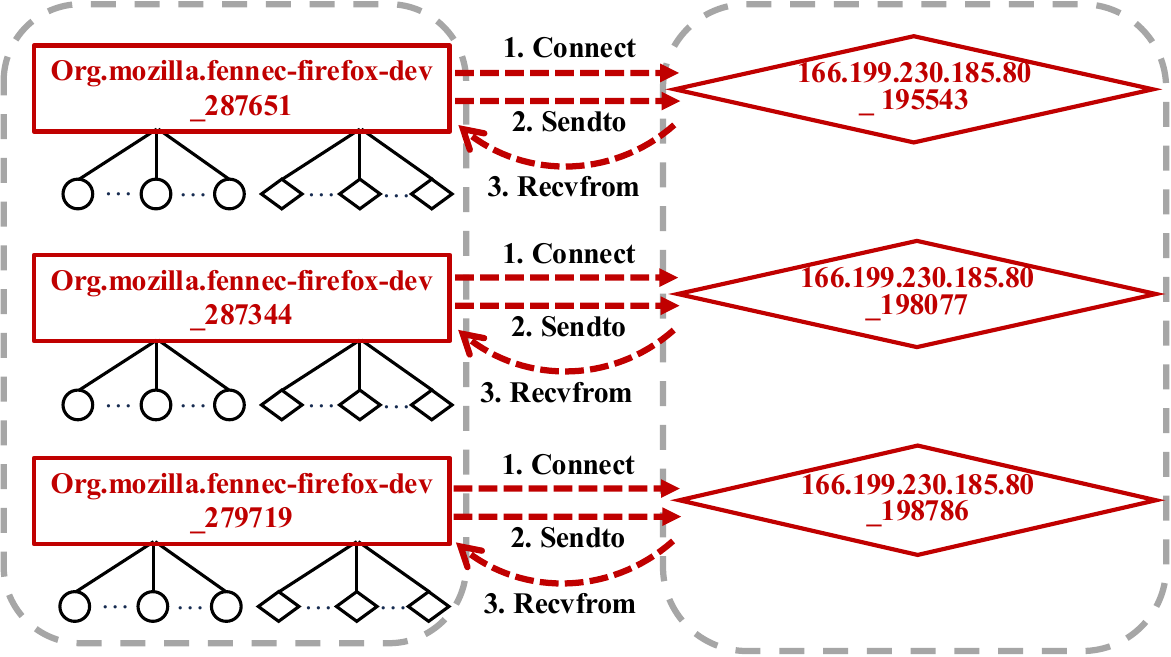}
\vspace{-5pt}
\caption{The attack subgraph generated by \system{} for the \texttt{CLEARSCOPE-E3} dataset, highlighting the key malicious events and their causal relationships. In this graph, circles represent \texttt{FILE}, squares represent \texttt{PROCESS}, and diamonds represent \texttt{NETWORK} nodes. The entity marked as red denotes true positives detected by \system{}. }
\vspace{-5pt}
\label{fig:case_study_graph}
\end{figure}

\rev{Beyond aggregate metrics, we examine whether \system{} produces \emph{actionable} alerts for analysts. We use the \texttt{CLEARSCOPE-E3} scenario because its attack surface is small: in our run, \system{} surfaced only \textbf{six} true-positive nodes. A 1-hop expansion around them yields a single concise subgraph (Figure~\ref{fig:case_study_graph}) centered on process \texttt{org.mozilla.fennec\_firefox\_dev} and its connection to \texttt{166.199.230.185:80}, already suggesting a plausible exfiltration path.}

\rev{\noindent\textbf{Per-Node Scoring Walkthrough.} Table~\ref{tab:walkthrough} traces the six true positives in Figure~\ref{fig:case_study_graph}, plus two false alarms for contrast, through raw view scores, percentile-normalized scores, and the final seven-detector vote vector.}

\begin{table}[h]
  \centering
  \small
  % \vspace{-5pt}
  {
  % \color{diffblue}
  \setlength{\tabcolsep}{3.5pt}
  \renewcommand{\arraystretch}{0.6}
  % \arrayrulecolor{diffblue}
  \begin{tabular}{llcccccccc}
  \toprule
   &  & \multicolumn{3}{c}{Raw} & \multicolumn{3}{c}{Norm} &  &  \\
  \cmidrule(lr){3-5}\cmidrule(lr){6-8}
   & UUID & $S_S$ & $S_A$ & $S_C$ & $S'_S$ & $S'_A$ & $S'_C$ & Votes & $V$ \\
  \midrule
  \multirow{6}{*}{\textit{TP}}
    & 287651 & 3.14 & 0.10 & 1.56 & 1.00 & 0.30 & 1.00 & \texttt{1011101} & 5 \\
    & 287344 & 1.39 & 0.10 & 1.59 & 1.00 & 0.30 & 1.00 & \texttt{1011101} & 5 \\
    & 279719 & 1.04 & 0.10 & 1.56 & 1.00 & 0.30 & 1.00 & \texttt{1011101} & 5 \\
    & 195543 & 0.92 & 0.41 & 1.56 & 1.00 & 1.00 & 1.00 & \texttt{0011111} & 5 \\
    & 198077 & 0.90 & 0.41 & 1.59 & 1.00 & 1.00 & 1.00 & \texttt{0011111} & 5 \\
    & 198786 & 0.80 & 0.41 & 1.56 & 0.99 & 1.00 & 1.00 & \texttt{0011111} & 5 \\
  \midrule
  \multirow{2}{*}{\textit{FP}}
    & 115182 & 0.84 & 0.05 & 1.57 & 1.00 & 0.09 & 1.00 & \texttt{1011101} & 5 \\
    & 284778 & 4.37 & 0.43 & 0.93 & 1.00 & 1.00 & 0.99 & \texttt{1101111} & 6 \\
  \bottomrule
  \end{tabular}
  }
  \vspace{-5pt}
  % \captionsetup{labelfont={color=diffblue}, textfont={color=diffblue}}
  \caption{Scoring trace. $S_S$/$S_A$/$S_C$ denote the attribute, structural, and causal view scores; primed symbols denote percentile-normalized scores. The vote vector lists $D_1$--$D_7$; an alert fires when $V \geq T_v = 4$.}
  \label{tab:walkthrough}
  \vspace{-3pt}
\end{table}

\rev{To keep decisions inspectable, \system{} exposes a per-entity vote vector. For \texttt{org.mozilla.fennec\_firefox\_dev}, the vector is \texttt{[1,0,1,1,1,0,1]} with normalized scores \([1.0, 0.3, 1.0]\), showing which internal detectors contributed to the alert. This allows analysts to verify cross-detector agreement (``5 of 7 voted anomalous'') and prioritize the associated edges and processes for follow-up.}

\rev{The resulting subgraph reveals a repeated pattern of \textit{file access} followed by \textit{network transmission}, observed three times, consistent with the ground truth that the adversary attempted exfiltration three times under unstable connectivity. Together, precise node selection and transparent voting provide a small, context-rich view that supports efficient attack reconstruction without producing many disconnected or low-information alerts.}

\begin{table}
\centering
\begin{small}
% \vspace{-8pt}
\begin{tblr}{
  rowsep = -0.5pt,
  colsep = 1.2pt,
  width = \linewidth,
  colspec = {Q[135]Q[113]Q[104]Q[113]Q[115]Q[104]Q[113]Q[113]},
  cells = {c},
  cell{1}{2} = {c=3}{},
  cell{1}{5} = {c=4}{},
  vline{2-3} = {1}{},
  vline{2,5} = {1-8}{},
  hline{1,9} = {-}{0.1em},
  hline{2} = {2-8}{l},
  hline{3,6} = {-}{0.05em},
}
                 & \textbf{Training Time (s)} &                     &                         & \textbf{Inference Time (s)} &                       &                      &                          \\
\textbf{Dataset} & \textbf{GMAE Train}     & \textbf{Causal MLP} & \textbf{Total Train} & \textbf{GNN Embed}      & \textbf{Causal Pred.} & \textbf{KNN Score} & \textbf{Total Infer} \\
CA-E3        & 9.00                       & 394.0              & 403.0                  & 1.3                        & 14.2                 & 14.8                & 30.3                    \\
TH-E3         & 4.5                       & 407.7              & 412.2                  & 1.3                        & 17.7                 & 15.0                & 33.0                    \\
CL-E3        & 4.0                       & 157.2              & 161.2                  & 0.4                        & 6.2                  & 3.2                 & 9.8                      \\
CA-E5        & 3190.5                    & 797.4              & 3987.8                 & 9.7                        & 86.4                 & 574.0              & 670.1                  \\
TH-E5         & 2100.1                    & 416.0              & 2516.1                 & 7.5                       & 72.9                & 169.0               & 249.4                   \\
CL-E5        & 118.5                     & 587.9              & 706.5                  & 0.97                        & 18.6                 & 5.0                 & 24.5                     
\end{tblr}
\vspace{-5pt}
\caption{Runtime breakdown for \system{}, highlighting the computational cost of each component.}
\vspace{-10pt}
\label{tab:runtime_breakdown}
\end{small}
\end{table}

\begin{table*}[t!]
\centering
\begin{small}
\begin{tblr}{
  rowsep = -0.5pt,
  colsep = 1.2pt,
  width = \linewidth,
  cell{2}{1} = {r=11}{},
  cell{13}{1} = {r=8}{},
  cell{21}{1} = {r=2}{},
  colspec = {Q[102]Q[277]Q[377]Q[181]},
  hline{1,23} = {-}{0.1em},
  hline{2,13,21} = {-}{0.05em},
}
Dataset & UUID & Attribute & Timestamp \\
THEIA-E3 &
ED35A9B7-0200-0000-0000-000020 & /home/admin/profile & 2018-04-12 12:57 \\
& 273847BC-0200-0000-0000-000020 & /home/admin/profile & 2018-04-12 12:57 \\
& D037D3BA-0200-0000-0000-000020 & /home/admin/profile~ & 2018-04-12 13:10 \\
& D237D6BA-0200-0000-0000-000020 & /home/admin/profile~ & 2018-04-12 13:10 \\
& EE35ABB7-0200-0000-0000-000020 & /home/admin/profile~ & 2018-04-12 12:57 \\
& 1D38E3BB-0200-0000-0000-000020 & /home/admin/profile~ & 2018-04-12 13:15 \\
& EC3598B7-0200-0000-0000-000020 & /home/admin/profile & 2018-04-12 12:57 \\
& 54387BBE-0200-0000-0000-000020 & /home/admin/profile & 2018-04-12 13:26 \\
& 223838BC-0200-0000-0000-000020 & /home/admin/profile~ & 2018-04-12 13:16 \\
& 62175519-0400-0000-0000-000020 & /bin/bash (./tcexec) & 2018-04-13 14:06 \\
& 0100D00F-2925-2E00-0000-1889CA01 & /tmp/mozilla\_admin0/jAG\_iSHt.bin.part~ & 2018-04-13 14:06 \\
Clearscope-E3 &
00000000-0000-0000-000028A44E & org.mozilla.fennec\_firefox\_dev & 2018-04-11 14:12 \\
& 00000000-0000-0000-00002740D6 & org.mozilla.fennec\_firefox\_dev & 2018-04-11 13:54 \\
& 00000000-0000-0000-0000279530 & 128.55.12.166 \textbar{} 45525~~~ \textbar{} 166.199.230.185 \textbar{} 80~ & 2018-04-11 13:54 \\
& 00000000-0000-0000-00002974E9 & 128.55.12.166 \textbar{} 46162~~~ \textbar{} 166.199.230.185 \textbar{} 80 & 2018-04-11 14:12 \\
& 00000000-0000-0000-00002B1ADE & 128.55.12.166 \textbar{} 46316~~~ \textbar{} 166.199.230.185 \textbar{} 80 & 2018-04-11 14:20 \\
& 00000000-0000-0000-000027871D~ & 128.55.12.166 \textbar{} 55331~~~ \textbar{} 111.82.111.27 \textbar{} 80 & 2018-04-11 13:54 \\
& 00000000-0000-0000-00002974D0 & 128.55.12.166 \textbar{} 55968~~~ \textbar{} 111.82.111.27 \textbar{} 80~~~~~~~ & 2018-04-11 14:11 \\
& 00000000-0000-0000-00002B1AB3 & 128.55.12.166 \textbar{} 56122~~~ \textbar{} 111.82.111.27 \textbar{} 80~~ & 2018-04-11 14:20 \\
Clearscope-E5 &
00000000-0000-0000-0000F3D696 & org.mozilla.fennec\_vagrant & 2019-05-17 11:57 \\
& 516DD5D2-8D47-CB75-4EC4-82FF9B5054BB & /data/data/org.mozilla.fennec\_vagrant/files/
mozilla/profiles.ini~ & 2019-05-17 11:57 \\
\end{tblr}
\vspace{-5pt}
\caption{The newly added True Positives.}
\vspace{-5pt}
\label{tab:ground_truth}
\end{small}
\end{table*}

\vspace{-8pt}
\appsection{The performance of Each Detector.}{apd:performance_of_each_detector}
\vspace{-10pt}
The standalone performance of each of the seven detectors is presented in Table~\ref{tab:detector_eac}. A key observation is that no single detector strikes an acceptable trade-off between detection rate and the number of false positives, which highlights the limitations of individual heuristics and underscores the need for an ensemble method. For example, the D6 detector, a simple fusion of three views, produces a high volume of false positives. Conversely, the D4 detector achieves greater precision but at the cost of more false alarms. These results confirm that any single method is inadequate on its own, providing the primary motivation for developing our ensemble model.

\begin{table*}[h!]
\centering
\begin{small}
\begin{tblr}{
  rowsep = -0.5pt,
  colsep = 2.4pt,
  cells = {c},
  width = 1.0\linewidth,
  % colspec = {Q[81]Q[69]Q[42]Q[58]Q[58]Q[42]Q[46]Q[52]Q[56]Q[69]Q[42]Q[42]Q[60]Q[42]Q[46]Q[52]Q[56]},
  cell{2}{2} = {r=7}{},
  cell{2}{10} = {r=7}{},
  cell{9}{2} = {r=7}{},
  cell{9}{10} = {r=7}{},
  cell{16}{2} = {r=7}{},
  cell{16}{10} = {r=7}{},
  hline{1,23} = {-}{0.1em},
  hline{2,9,16} = {-}{0.05em},
  vline{2,3,10,11} = {-}{0.05em},
}
System     & Dataset        & TP↑ & FP↓    & TN↑    & FN↓ & F1↑ & ADP↑ & MCC↑ & Dataset        & TP↑ & FP↓    & TN↑    & FN↓ & F1↑ & ADP↑ & MCC↑ \\
Kairos     & {CADETS \\ E3} & 1   & 959    & 280.6K & 67  & 0.00 & 0.00 & 0.00 & {CADETS \\ E5} & 0   & 6      & 3.14M  & 123 & 0    & 0.01 & 0    \\
Magic      &                & 22  & 16.5K  & 265.0K & 46  & 0.00 & 0.01 & 0.02 &                & 28  & 245.3K & 2.89M  & 95  & 0.00 & 0.17 & 0.00 \\
NodLink    &                & 18  & 34.3K  & 247.2K & 50  & 0.00 & 0.49 & 0.01 &                & 73  & 756.0K & 2.38M  & 50  & 0.00 & 0.03 & 0.01 \\
Flash      &                & 3   & 4.5K   & 277.0K & 65  & 0.00 & 0.04 & 0.01 &                & 6   & 34961  & 3.10M  & 117 & 0.00 & 0.02 & 0.01 \\
Orthrus    &                & 7   & 1      & 281.5K & 61  & 0.18 & 0.81 & 0.30 &                & 1   & 8      & 3.14M  & 122 & 0.00 & 0.34 & 0.03 \\
Velox      &                & 9   & 1      & 281.5K & 59  & 0.23 & 0.97 & 0.35 &                & 0   & 2      & 3.14M  & 123 & 0.00 & 0.01 & 0.00 \\
ProvFusion &                & 24  & 1      & 281.5K & 44  & 0.52 & 1.00 & 0.58 &                & 7   & 9      & 3.14M  & 116 & 0.10 & 0.92 & 0.16 \\
Kairos     & {THEIA \\ E3}  & 1   & 22     & 701.4K & 117 & 0.01 & 0.11 & 0.02 & {THEIA \\ E5}  & 0   & 7      & 1.86M  & 69  & 0.00 & 0.00 & 0.00 \\
Magic      &                & 19  & 97.2K  & 604.2K & 99  & 0.00 & 0.00 & 0.00 &                & 61  & 737.3K & 1.12M  & 8   & 0.00 & 0.00 & 0.00 \\
NodLink    &                & 26  & 258.3K & 443.2K & 92  & 0.00 & 0.01 & 0.00 &                & 20  & 175.3K & 1.68M  & 49  & 0.00 & 0.00 & 0.00 \\
Flash      &                & 20  & 251.2K & 450.2K & 98  & 0.00 & 0.01 & 0.00 &                & 41  & 316.2K & 1.54M  & 28  & 0.00 & 0.01 & 0.01 \\
Orthrus    &                & 2   & 2      & 701.4K & 116 & 0.03 & 0.33 & 0.09 &                & 1   & 31     & 1.86M  & 68  & 0.02 & 0.30 & 0.02 \\
Velox      &                & 18  & 120    & 701.3K & 100 & 0.14 & 0.89 & 0.14 &                & 2   & 63     & 1.86M  & 67  & 0.03 & 0.33 & 0.03 \\
ProvFusion &                & 82  & 11     & 701.4K & 36  & 0.78 & 1.00 & 0.78 &                & 11  & 2      & 1.86M  & 58  & 0.27 & 1.00 & 0.37 \\
Kairos     & {CLRSCP \\ E3} & 6   & 8.4K   & 102.9K & 35  & 0.00 & 0.00 & 0.01 & {CLRSCP \\ E5} & 1   & 1      & 150.9K & 50  & 0.04 & 0.37 & 0.10 \\
Magic      &                & 37  & 8.4K   & 102.9K & 4   & 0.01 & 0.01 & 0.06 &                & 7   & 8.2K   & 142.6K & 44  & 0.00 & 0.00 & 0.01 \\
NodLink    &                & 38  & 22.8K  & 88.5K  & 3   & 0.00 & 0.06 & 0.03 &                & 3   & 27.0K  & 123.9K & 48  & 0.00 & 0.00 & 0.00 \\
Flash      &                & 32  & 11.1K  & 100.2K & 9   & 0.01 & 0.03 & 0.04 &                & 17  & 76.0K  & 74.9K  & 34  & 0.00 & 0.01 & 0.00 \\
Orthrus    &                & 1   & 9      & 111.3K & 40  & 0.00 & 0.17 & 0.05 &                & 2   & 8      & 150.9K & 49  & 0.07 & 0.21 & 0.09 \\
Velox      &                & 1   & 625    & 110.7K & 40  & 0.00 & 0.10 & 0.00 &                & 8   & 13     & 150.9K & 43  & 0.22 & 0.50 & 0.24 \\
ProvFusion &                & 1   & 12     & 111.3K & 40  & 0.04 & 0.50 & 0.04 &                & 9   & 17     & 150.9K & 42  & 0.23 & 0.71 & 0.25 
\end{tblr}
\vspace{-5pt}
\caption{Results under the Original Orthrus Ground Truth}
\vspace{-10pt}

\label{tab:orthrus_results}
\end{small}
\end{table*}

\vspace{-8pt}
\appsection{Added New Ground-Truth Entities.}{apd:new_ground_truth}
\vspace{-10pt}
As illustrated in Section~\S\ref{sec:node_level_performance}, we found and added new true positives into the ground truth file. Here we proved a Table~\ref{tab:ground_truth} that details the UUID, attribute, and timestamp for each added entity.

%=======================================================================
\vspace{-8pt}
\appsection{Detailed Runtime Performance Breakdown}{app:runtime_breakdown}
\vspace{-10pt}
%=======================================================================

% 1. 引言：解释这个附录的目的
To support our analysis of computational overhead in Section~\ref{sec:overhead}, this section provides a detailed breakdown of the runtime components for \system{} across all six datasets. All measurements were conducted on the same hardware (NVIDIA A100 GPU) and averaged over five runs.

Table~\ref{tab:runtime_breakdown} provides definitive empirical evidence for our claim in Section~\ref{sec:overhead}. 

\vspace{-8pt}
\appsection{Mimicry Attack Evaluation}{sec:appendix_robustness}
\vspace{-10pt}

We reproduced the mimicry attack of Goyal et al.~\cite{sometimes} to assess whether subgraph-level perturbations affect \system{}. The attacker inserted 1K–5K camouflage edges into malicious subgraphs to imitate benign motifs. As shown in Table~\ref{tab:mimicry}, all original malicious entities remained correctly detected, while false positives slightly increased with higher perturbation levels. Overall, mimicry perturbations reduced MCC modestly but did not conceal true malicious behaviors, indicating that localized detectors in node/edge-centric paradigms remain resilient to subgraph-level manipulations.

\begin{table}[h]
% \vspace{-5pt}
\centering
\setlength{\tabcolsep}{5pt} % 调整列间距 (默认约6pt)
\small % 或 \footnotesize 根据空间需求
\begin{tblr}{
  width = 0.6\linewidth,
  rowsep = 0pt,
  colsep = 1.2pt,
  colspec = {Q[475]Q[133]Q[142]Q[142]},
  column{even} = {c},
  column{3} = {c},
  hline{1,6} = {-}{0.1em},
  hline{2} = {-}{0.05em},
}
\textbf{Added Edges} & \textbf{1000} & \textbf{3000} & \textbf{5000} \\
True Positives (TP)  & 24            & 23            & 22            \\
False Positives (FP) & 9             & 31            & 34            \\
MCC                  & 0.51          & 0.38          & 0.36          \\
ADP                  & 0.92          & 1.00          & 1.00          
\end{tblr}
\vspace{-5pt}
\caption{Performance of \system{} under mimicry attacks with different numbers of added edges.}
\vspace{-10pt}
\label{tab:mimicry}
\end{table}

\vspace{-8pt}
\appsection{Results under the Orthrus Ground Truth}{apd:orthrus_results}
\vspace{-10pt}

We repeated all evaluations using the original Orthrus ground truth to verify consistency  (Table~\ref{tab:orthrus_results}). The overall performance trends of \system{} and all baselines remain unchanged, confirming that the additional ground-truth refinements do not alter the relative ranking or main conclusions. Minor variations in absolute metrics are observed, primarily due to the previously missing attacks.

\vspace{-8pt}
\appsection{Hyperparameter Settings for Comparative Anomaly Detection Algorithms}{apd:parameterexplore}
\vspace{-10pt}

Here we summarize the hyperparameter ranges explored for each model family mentioned in Section \S\ref{sec:fusion_rationale}.
All settings were applied under identical data partitions and four normalization schemes (min-max, z-score, robust, and percentile) for strict comparability.

\noindent\textbf{Linear Combination Models.}
All integer weight triplets $(w_{\text{attr}},w_{\text{struc}},w_{\text{causal}})!\in!{0,\ldots,19}^3!\setminus!{(0,0,0)}$ were enumerated, yielding 7,999 combinations. 

\noindent\textbf{Probabilistic Models.}
\textit{Multivariate Gaussian (MVG):} regularization coefficient added to the covariance diagonal
$\text{regularization}_{\text{cov}}\in{10^{-2},10^{-3},10^{-4},10^{-5},10^{-6}}$.
\textit{Gaussian Mixture Model (GMM):} number of mixture components
$n{\text{components}}\in{1,3,5,7,10,12,15}$. 

\noindent\textbf{Distance/Density-based Models.}
\textit{K-Nearest Neighbor (KNN):} neighborhood size
$n_{\text{neighbors}}\in{5,10,20,30,40,50}$. 

\noindent\textbf{Boundary-based Models.}
\textit{One-Class SVM (OC-SVM):} error upper bound parameter $\nu\in{10^{-3}}$, $10^{-4}$, $10^{-5}$, $10^{-6}$, $10^{-7}$.
\textit{Isolation Forest (iForest):} number of estimators
$n_{\text{estimators}}\in{50,100,150,200,300,500,700,900}$. 

\noindent\textbf{Reconstruction-based Model.}
\textit{Autoencoder:} learning rate
$\text{lr}\in{10^{-2},10^{-3},10^{-4},10^{-5},10^{-6}}$.

The best-performing configuration per dataset was reported in the main results (Table~\ref{tab:scorer_comparison}).

 \rev{
  \vspace{-7pt}
\section{Multi-Hot vs.\ Per-Edge Encoding}
 \vspace{-8pt}
  \label{sec:multihot_ablation}
  To verify that collapsing parallel events between the same entity pair
  into a single multi-hot edge does not degrade detection, we compare
  both encodings on THEIA-E3 under identical settings. We choose
  THEIA-E3 because its larger pool of true positives makes any
  encoding-induced differences clearly observable.}

\begin{table}[h]
  \vspace{-5pt}
  \centering
  \small
  {
  % \color{diffblue}
  \setlength{\tabcolsep}{0pt}
  \setlength{\tabcolsep}{5pt}
  % \arrayrulecolor{diffblue}
  \begin{tabular}{lccccccc}
  \toprule
  Encoding & TP & FP & TN & FN & F1 & ADP & MCC \\
  \midrule
  Multi-hot (Ours) & 91  & 2  & 701K & 38 & 0.82 & 1.00 & 0.83 \\
  Per-edge         & 108 & 19 & 701K & 21 & 0.84 & 1.00 & 0.84 \\
  \bottomrule
  \end{tabular}
  }
  \vspace{-5pt}
  % \captionsetup{labelfont={color=diffblue}, textfont={color=diffblue}}
  \caption{Multi-hot vs. per-edge encoding on THEIA-E3.}
    \vspace{-7pt}
  \label{tab:encoding_compare}
\end{table}

\rev{  The two encodings achieve essentially equivalent MCC; per-edge recovers
  a few extra TPs but at nearly $10\times$ the FP cost, confirming that
  multi-hot is a sound efficiency-motivated choice rather than a detection
  compromise.}
% \appendices % if not used earlier
\newpage

\rev{

\appsection{Meta-Review}{apd:meta-review}

The following meta-review was prepared by the program committee for the 2026
IEEE Symposium on Security and Privacy (S\&P) as part of the review process as
detailed in the call for papers.

\subsection{Summary}
\system{} is a multi-view provenance-based intrusion detection framework that fuses anomaly signals from three views (attribute, structure, causality) using percentile normalization, monotonic linear detectors, and voting-based aggregation. It is evaluated on DARPA TC datasets and outperforms single-view baselines in detection accuracy and false-positive rates.

\subsection{Scientific Contributions}
\begin{itemize}
\item Creates a New Tool to Enable Future Science.
\item Provides a Valuable Step Forward in an Established Field.
\end{itemize}

\subsection{Reasons for Acceptance}
\begin{enumerate}
\item Well-motivated problem with a compelling diagnosis of why single-view detectors fail in complementary ways.
\item Clear empirical gains over both individual baselines and naive ensemble combinations, validated through extensive ablations.

\item The system is lightweight and practical. The fusion and voting mechanism adds minimal overhead, and the evaluation shows improvements in runtime efficiency.
\end{enumerate}

\subsection{Noteworthy Concerns} % Exclude if your meta-review does not have noteworthy concerns
\begin{enumerate} % Enumerate environment is not necessary if there is only one
\item Reviewers flagged an area where finer detail would strengthen the paper: a per-campaign FP/FN breakdown.
\end{enumerate}
}
\end{document}